\documentclass[twocolumn]{aastex631}
\usepackage [latin1]{inputenc}

\usepackage{natbib}
\usepackage{amsmath}
\usepackage{graphicx}
\usepackage{multirow}

\newcommand{\lib}[1]{\texttt{#1}}
\newcommand{\pkg}[1]{\texttt{#1}}
\newcommand{\alphain}{\alpha_{\text{in}}}
\newcommand{\alphaout}{\alpha_{\text{out}}}
\newcommand{\amin}{a_{\text{min}}}
\newcommand{\amax}{a_{\text{max}}}

\newcounter{savefootnote}
\newcounter{symfootnote}
\newcommand{\symfootnote}[1]{%
   \setcounter{savefootnote}{\value{footnote}}%
   \setcounter{footnote}{\value{symfootnote}}%
   \ifnum\value{footnote}>8\setcounter{footnote}{0}\fi%
   \let\oldthefootnote=\thefootnote%
   \renewcommand{\thefootnote}{\fnsymbol{footnote}}%
   \footnote{#1}%
   \let\thefootnote=\oldthefootnote%
   \setcounter{symfootnote}{\value{footnote}}%
   \setcounter{footnote}{\value{savefootnote}}%
}

\newcommand{\farcsec}{\hbox{$.\!\!^{\prime\prime}$}}

\newcommand{\hd}{HD~141569A}

\renewcommand{\bfseries}{\mdseries}

\begin{document}

\title{Probing Disk Ice Content and PAH Emission Through Multiband MagAO+Clio Images of HD~141569}

\shorttitle{Multiband MagAO+Clio Imaging of HD 141569}
\shortauthors{Kueny et al.}

\author[0000-0001-8531-038X]{Jay K. Kueny}
\affiliation{Steward Observatory, University of Arizona, Tucson, 933 N Cherry Ave, Tucson, AZ 85721, USA}
\affiliation{James C. Wyant College of Optical Sciences, University of Arizona, 1630 E. University Blvd., Tucson, AZ 85721, USA}
\affiliation{National Science Foundation Graduate Research Fellow}

\author[0000-0001-6654-7859]{Alycia J. Weinberger}
\affiliation{Earth and Planets Laboratory, Carnegie Institution for Science, 5241 Broad Branch Road NW, Washington, DC 20015-1305}

\author[0000-0002-2346-3441]{Jared R. Males}
\affiliation{Steward Observatory, University of Arizona, Tucson, 933 N Cherry Ave, Tucson, AZ 85721, USA}

\author[0000-0002-1384-0063]{Katie M. Morzinski}
\affiliation{Steward Observatory, University of Arizona, Tucson, 933 N Cherry Ave, Tucson, AZ 85721, USA}

\author[0000-0002-2167-8246]{Laird M. Close}
\affiliation{Steward Observatory, University of Arizona, Tucson, 933 N Cherry Ave, Tucson, AZ 85721, USA}

\author[0000-0002-7821-0695]{Katherine B. Follette}
\affiliation{Department of Physics and Astronomy, Amherst College, Amherst, MA 01003, USA}

\author[0000-0002-1954-4564]{Philip M. Hinz}
\affiliation{UC Santa Cruz, 1156 High St, Santa Cruz CA 95064, USA}

\begin{abstract}

We present resolved images of the inner disk component around \hd{} using the Magellan adaptive optics system with the Clio2 1 - $5 \mu$m camera, offering a glimpse of a complex system thought to be in a short evolutionary phase between protoplanetary and debris disk stages. We use a reference star along with the KLIP algorithm for PSF subtraction to detect the disk inward to about $0\farcsec24\ (\sim25$ au assuming a distance of 111 pc) at high signal-to-noise ratios at $L'\ (3.8 \mu$m), $Ls\ (3.3 \mu$m), and narrowband $Ice\ (3.1 \mu$m). We identify an arc or spiral arm structure at the southeast extremity, consistent with  previous studies. We implement forward modeling with a simple disk model within the framework of an MCMC sampler to better constrain the geometrical attributes and photometry using our KLIP-reduced disk images. We then leverage these modeling results to facilitate a comparison of the measured brightness in each passband to find a reduction in scattered light from the disk in the $Ice$ filter, implying significant absorption due to water ice in the dust. Additionally, our best-fit disk models exhibit peak brightness in the southwestern, back-scattering region of the disk, which we suggest to be possible evidence of 3.3 $\mu$m PAH emission. However, we point out the need for additional observations with bluer filters and more complex modeling to confirm these hypotheses.

\end{abstract}


\section{Introduction} 
\label{sec:intro}

Circumstellar disks offer unique opportunities to study  how planets form and interact with their environments. Observations of disks reveal a broad range of developmental periods, ranging from young, gas-rich protoplanetary disks (e.g., HL Tauri; \citealt{alma_partnership_2014_2015}) to gas-depleted systems showcasing complex debris rings akin to the Asteroid Main Belt (e.g., Formalhaut; \citealt{gaspar_spatially_2023}). It is the transitional and ``hybrid disk" stages that are of particular value as a laboratory in which to study the last stages of gas-giant and collisional stages of terrestrial planet formation. \textbf{Young to intermediate-age disks ($\lesssim 10$ Myr) such as the HD 141569 system can thus help refine these theories with direct imaging of nascent planets or by observing characteristic features of early-stage planet formation such as spiral arms, gaps, and asymmetrical dust distribution \citep{wyatt_spiral_2005}.}

The distribution of bulk compositions including water, carbon, and silicon in circumstellar disks may offer clues to the ultimate composition and evolution of planets \citep{shahar_what_2019}. One way to ascertain the presence of water ice in disks is to observe the scattered light at 3--4 $\mu$m including the water-ice absorption feature at $3.1 \mu$m \citep{inoue_observational_2008}.  This method has revealed water ice in the scattering surface layers of the disks around HD 142527 \citep{honda_detection_2009}, HD 100546 \citep{honda_water_2016}, and AB Aurigae \citep{betti_detection_2022}.

Polycyclic aromatic hydrocarbons (PAHs) are an easily detected form of carbon in the disks around luminous stars. In response to ultraviolet (UV) excitation, these large molecules display distinct emission features in the 3-$20 \mu$m spectral range; see the review by \cite{tielens_interstellar_2008} for a thorough discussion. Young disks are likely to be permeated with PAHs, but the uppermost disk layers receive the highest levels of UV flux and therefore have the highest PAH excitation rates \citep{siebenmorgen_polycyclic_2012}. Because of this, it has been proposed \citep{seok_polycyclic_2017} that PAH emission can be a tracer for the gas-dominated disk atmosphere. Furthermore, given their stochastic heating mechanism \citep{draine_infrared_2001}, PAH emission may be easily detected at large radial distances from the star on both the front and back sides, as long as we have a direct line-of-sight to the emitters. PAHs substantially impact circumstellar disk chemistry in the later evolutionary stages, acting as the formation site of, e.g., molecular hydrogen and water \citep{jonkheid_modeling_2006}. 

\hd{} is a young ($5 \pm 3$ Myr; \citealt{weinberger_circumstellar_1999}) Herbig B9.5V/A0V star surrounded by a complex multi-ring system featuring gaps, spirals, arcs, and material asymmetries (see, e.g., \citealt{augereau_hstnicmos2_1999}; \citealt{weinberger_circumstellar_1999}; \citealt{mouillet_asymmetries_2001}; \citealt{clampin_hubble_2003}; \citealt{konishi_discovery_2016}; \citealt{mawet_characterization_2017}; \citealt{di_folco_almanoema_2020}; \citealt{bruzzone_imaging_2020}; \citealt{singh_revealing_2021}). \textbf{This unique disk is categorized as ``hybrid" (\citealt{wyatt_five_2015}; \citealt{di_folco_almanoema_2020}) in that it has substantial molecular gas \citep[e.g.,][]{zuckerman_inhibition_1995} and PAHs yet also displays the low optical depth of an early-stage debris disk.} It may be in the final phases of dissipation and the last gasps of giant planet formation and migration. The host star is not variable \citep{alvarez_thriteen-color_1981}\textbf{, is of intermediate mass with a luminosity} of $27.5 \pm 0.95 L_{\odot}$ (calculated from the flux reported in \citealt{merin_study_2004} and the distance of $111.6 \pm 04$ pc reported in Gaia Data Release 3; \citealt{gaia_collaboration_gaia_2023}). Two M dwarf stars are located about 8 arcseconds from \hd{} and may be bound companions \citep{weinberger_stellar_2000}, or have recently passed within 900 au \citep{reche_investigating_2009}. Gravitational effects either from these two low-mass companions \citep{augereau_structuring_2004,reche_investigating_2009} and/or from an unresolved giant planet \citep{wyatt_spiral_2005} could be responsible for spiral arms seen in the disk.

The inner disk around \hd{} ($R \leq 90$ au) has \textbf{been the subject of multiple studies}, but the results of many of these analyses raised further questions. Direct imaging at near-infrared wavelengths using VLT/SPHERE (\citealt{perrot_discovery_2016}; \citealt{singh_revealing_2021}) revealed a complex network of concentric rings from 40 to 90 au. \cite{currie_matryoshka_2016} reported additional independent detections of complex features in the $L'$ band at $\sim40$ au using Keck/NIRC2. However, it is important to note that these imaging efforts made use of angular differential imaging (ADI); thus these images are, due to the extended geometry and moderate inclination of the disk, inherently affected by observational bias \citep{milli_impact_2012}. One way to mitigate this issue is using reference star differential imaging (RDI, \citealt{soummer_five_2014}; \citealt{ruane_reference_2019}) to carry out the point-spread function (PSF) subtraction. Indeed, RDI with the Keck/NIRC2 vortex coronagraph mode by \cite{mawet_characterization_2017} revealed a more radially-extended $L'$ profile from $\sim20$ to 70 au that is distinctly different than the aforementioned ADI images in colocated regions. \cite{mawet_characterization_2017} also revealed the presence of a spiral or arc-like feature extended across the outer boundary of the southeast region of the disk, perhaps tying into more recent GPI $H$-band polarimetry direct imaging results by \cite{bruzzone_imaging_2020} who also identified a separate, but proximal spiral or arc candidate in the south disk ansa. \textbf{A common conclusion in these studies (\citealt{perrot_discovery_2016}; \citealt{mawet_characterization_2017}; \citealt{bruzzone_imaging_2020}; \citealt{singh_revealing_2021}) is a significant brightness enhancement at the south ansa of the disk indicating a higher dust density distribution since the disk is optically-thin \citep{thi_gas_2014}.}

Gas is co-located with the dust disk. \cite{thi_gas_2014} found $8.6 \mu$m emission using VLT/VISIR which was thought to be from PAHs. Millimeter-wave range observations using ALMA/NOEMA unveiled significant asymmetry of $^{12}$CO $J = 3 \rightarrow 2$ emission in the western half of the SPHERE rings, possibly suggesting dynamical interactions between dust and gas in that region \citep{di_folco_almanoema_2020}. Finally, recent observations using VLTI/GRAVITY revealed an additional ring of likely dusty material located at $\sim1$ au \citep{gravity_collaboration_gravity_2021}.

In this work, we present imaging and forward modeling of the inner disk around \hd{},  which imply the detection of water ice and reveal the spatial distribution of $3.3 \mu$m PAH emission. We introduce the observations and detail the data reduction in the next section. In Section \ref{sec:modeling}, we discuss our modeling procedures. We present the results of our forward modeling in Section \ref{sec:results} and discuss the most likely scenarios to explain our water ice and PAH emission hypotheses in Section \ref{sec:discussion}. We use Section \ref{sec:summary} to conclude and summarize our results as well as encourage avenues for future work.

\section{Observations and Data Reduction} \label{sec:obs-title}

\subsection{Observations}
\label{sec:obs}

\begin{table}
\centering
\caption{\hd{} and Reference Star Properties}
\begin{tabular}{ccc}
\hline
 & HD 141569A           & HD 144271 \\ \hline
R.A. (J2000)    & 15 49 57.75     &   16 05 8.49   \\
Decl. (J2000)   & -03 55 16.34    &   -3 31 39.40 \\
Spectral Type   & B9.5V/A0Ve      &      A0     \\
$K$ (mag)       & $6.82 \pm 0.03$ &       $6.23 \pm 0.02$    \\
$L$ (mag)       & $6.06 \pm 0.07$ &      $6.21 \pm 0.05$     \\ \hline
\end{tabular}
\label{tab:properties}
\end{table}

\begin{table*}
\centering
\caption{Clio Observation Log of HD 141569A and PSF Reference}
\begin{tabular}{lclccccccc}
\hline
Date        & Time Start/End (UT)    & Object     & Band  & Air Mass & $t_{\text{exp}}$ (s) & $N_{\text{coadds}}$ & $N_{\text{exp}}$ & $N_{\text{nods}}$ & $\phi (^{\circ})$ \\ \hline
2014 Apr 10 & 06:51/07:47 & HD 141569A & $L'$  & 1.1/1.1  & 1                    & 20                  & 121              & 3                 & -15/14            \\
2014 Apr 10 & 07:54/09:15 & HD 144271  & $L'$  & 1.1/1.1  & 0.75                 & 20                  & 80               & 2                 & 9.5/41            \\
2014 Apr 10 & 08:12/08:57 & HD 141569A & $L'$  & 1.1/1.2  & 0.75                 & 20                  & 118              & 3                 & 25/40             \\
2014 Apr 11 & 05:58/06:39 & HD 141569A & $Ls$  & 1.1/1.1  & 2                    & 10                  & 90               & 2                 & -34/-19           \\
2014 Apr 11 & 06:44/06:52 & HD 144271  & $Ls$  & 1.1/1.1  & 2                    & 10                  & 20               & 1                 & -22/-19           \\
2014 Apr 11 & 06:56/07:19 & HD 141569A & $Ls$  & 1.1/1.1  & 2                    & 10                  & 58               & 1.5               & -10/2             \\
2014 Apr 12 & 05:22/05:31 & HD 141569A & $Ls$  & 1.2/1.2  & 2.5                  & 10                  & 60               & 3                 & -43/-41           \\
2014 Apr 12 & 05:58/06:03 & HD 144271  & $Ls$  & 1.1/1.1  & 2.5                  & 5                   & 20               & 1                 & -37/-36           \\
2014 Apr 12 & 06:06/06:16 & HD 141569A & $Ls$  & 1.1/1.1  & 2.5                  & 10                  & 20               & 1                 & -30/-27           \\
2014 Apr 12 & 07:22/07:31 & HD 141569A & $Ls$  & 1.1/1.1  & 2.5                  & 10                  & 20               & 1                 & 5/10              \\
2014 Apr 12 & 08:17/08:21 & HD 144271  & $Ls$  & 1.1/1.1  & 2.5                  & 4                   & 20               & 1                 & 24/25             \\
2015 May 30 & 02:46/03:17 & HD 141569A & $Ice$ & 1.2/1.1  & 14                   & 3                   & 43               & 2                 & -35/-23           \\
2015 May 30 & 03:22/03:51 & HD 144271  & $Ice$ & 1.1/1.1  & 14                   & 3                   & 40               & 2                 & -27/-14           \\
2015 May 30 & 03:55/04:52 & HD 141569A & $Ice$ & 1.1/1.1  & 10                   & 2                   & 160              & 4                 & -5/23             \\
2015 May 30 & 04:55/05:09 & HD 144271  & $Ice$ & 1.1/1.1  & 10                   & 2                   & 42               & 1                 & 17/24             \\
\hline
            &              &            &       &          &                      &                     &                  &                   &                  
\end{tabular}

\footnotesize
\raggedright
\textbf{Notes.} $t_{\text{exp}}$: single frame exposure time; $N_{\text{coadds}}$: number of coadded frames per image; $N_{\text{exp}}$: number of images; $N_{\text{nods}}$: number of ABBA nod sequences; $\phi$: start/end parallactic angle.
\label{tab:obslog}
\end{table*}

We obtained non-coronagraphic diffraction-limited images of \hd{} (spectral type B9.5-A0, K=6.8 mag, L=6.1 mag) and PSF reference HD 144271 (spectral type A0, K=6.2 mag, L=6.2 mag) using the 6.5 m Magellan Clay telescope at Las Campanas Observatory (LCO) on the following nights: UT 2014 April 9 - 11, UT 2015 May 28 and 29, and UT 2018 April 28. for reference, we include useful properties of our targets in Table \ref{tab:properties}. For all observations, we used the Magellan Adaptive Optics system (MagAO; \citealt{morzinski_magao_2016}) with the $1-5 \mu$m Clio-2 instrument in imaging mode (Clio hereafter; \citealt{morzinski_magellan_2015}) configured with its narrow camera that provided a plate scale of $0\farcsec01585$ per pixel and $16" \times 18"$ field of view. To facilitate RDI, we observed our target and reference stars with the rotator off to stabilize the pupil with respect to the detector focal plane. We performed ABBA nod sequences for our target and reference stars where the telescope was nodded by several arcseconds and dithered by a few tenths of an arcsecond for sky background and bad pixel removal during post-processing. To enable deeper disk images, we selected exposure times to saturate the detector out to about $0\farcsec07$ from the star. At the end of each nod sequence, we collected photometric calibration exposures (i.e., unsaturated images). We logged observational circumstances for each dataset that produced a final reduced image of acceptable quality in Table \ref{tab:obslog} (some observations were carried out in poor weather, see below).

We observed the target and reference using the $L'$ $(3.8 \mu$m) filter on the nights of UT 2014 April 9, UT 2015 May 28, and UT 2018 April 28. Conditions on the first night were initially excellent with $0\farcsec5$ seeing, but the seeing became variable during the latter half of this dataset prompting us to exclude the last 118 science frames (out of 240 total). Conditions on UT 2015 May 28 and UT 2018 April 28 were cloudy, with extinction affecting image quality and AO performance significantly so we ignored data from these nights in our analysis.

We collected images in the $3.3 \mu$m filter ($Ls$ for ``L-short") on the nights of UT 2014 April 10 and 11. Technical issues with the AO system along with cloudy conditions resulted in images limited in both quality and quantity for the first night. Conditions for UT 2014 April 11 were much improved and photometric, with seeing ranging from $0\farcsec5$ to $0\farcsec75$.

We imaged through a narrowband $Ice$ $(3.1 \mu$m) filter on the nights of UT 2014 April 11 and UT 2015 May 29. The seeing on the second night was quite good at $0\farcsec6$ - $0\farcsec7$ and very stable. \textbf{The internal optics of Clio create an illumination artifact in the center of the detector \citep{morzinski_magellan_2015}. Both the target and reference stars were imaged close to the center of the detector by mistake. This unfortunately caused an optical ghost to propagate through to the final image which manifests as a circular oversubtracted region in the final reduced disk image just east of the star.} The final reduced image using data from the first night was comparatively of lesser quality perhaps due to a moderate amount of sky rotation over the dataset ($\sim39^{\circ}$). \textbf{Because of the residual Airy pattern left in the final image, preliminary modeling using the data from UT 2014 April 11 produced models with non-physical parameters}, so we refrained from using these data in our study.

Additionally, for the night of UT 2014 April 11, we obtained data at $2.15 \mu$m ($Ks$ for ``K-short''). Because of time lost to an AO problem, the amount of sky rotation was only $\sim35^{\circ}$. We obtained only a small library of 16 reference PSF images for the RDI reduction (compared to 95 target images). Additionally, Clio's performance at $Ks$ is limited due to detector read noise. We did not detect any disk components in these data.

As a result of these various issues, we only included the $L'$ data taken on UT 2014 April 9, the $Ls$ data taken on UT 2014 April 12, and the $Ice$ data taken on UT 2015 May 29 in our modeling efforts (see Section \ref{sec:modeling}). For each of these three passbands we treated the libraries of target and reference star images as separate datasets and applied all data reduction steps mentioned below consistently to each.

\begin{figure*}[h!]
    \centering
    \includegraphics[width=\linewidth]{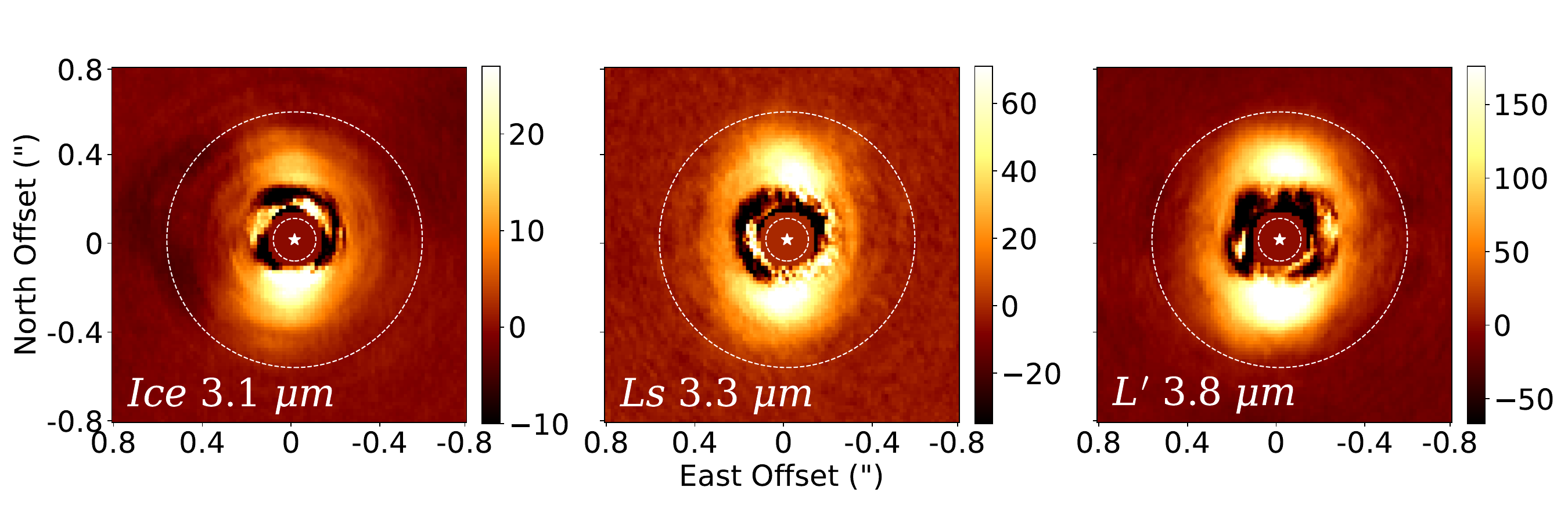} \\
    \includegraphics[width=0.32\linewidth]{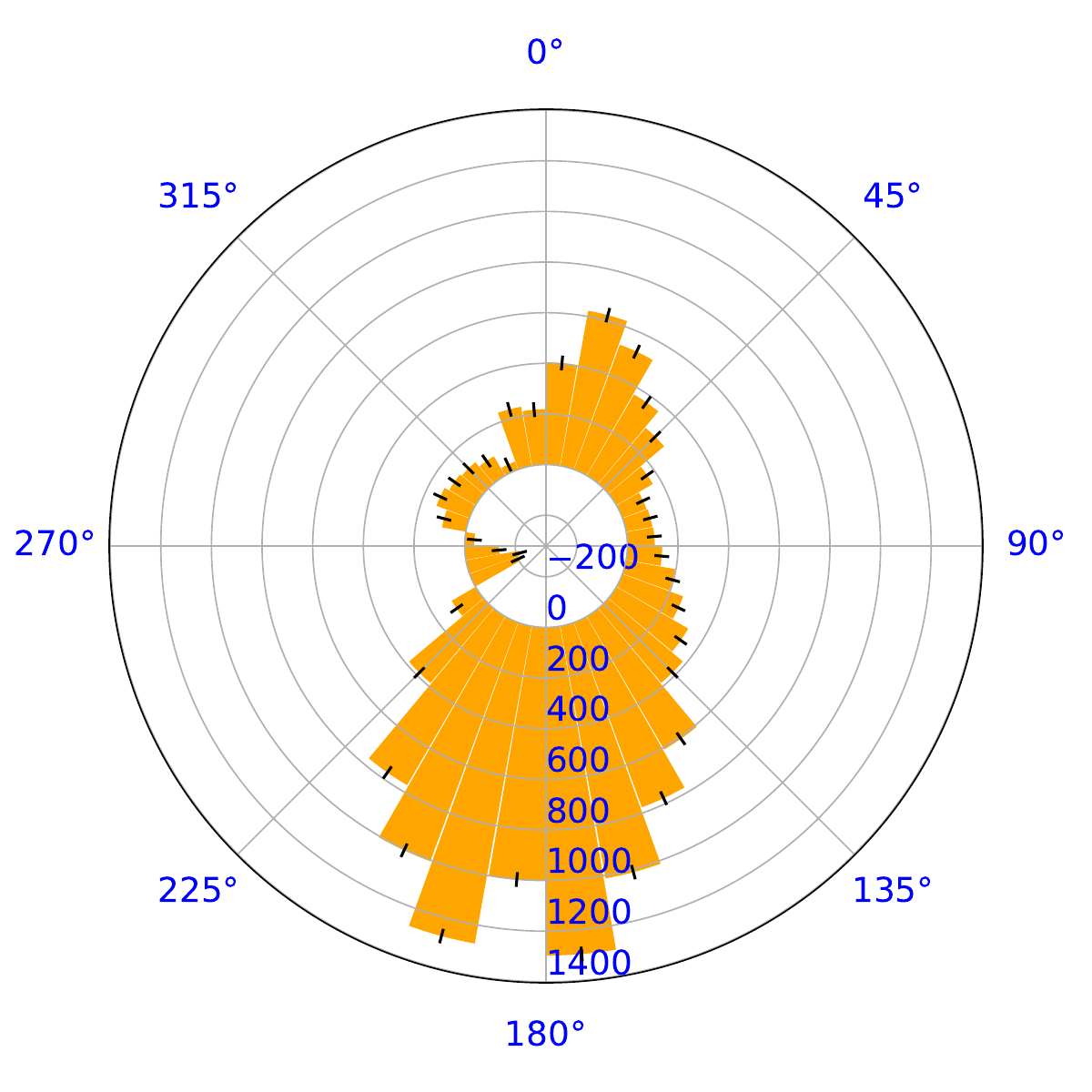} 
    \includegraphics[width=0.32\linewidth]{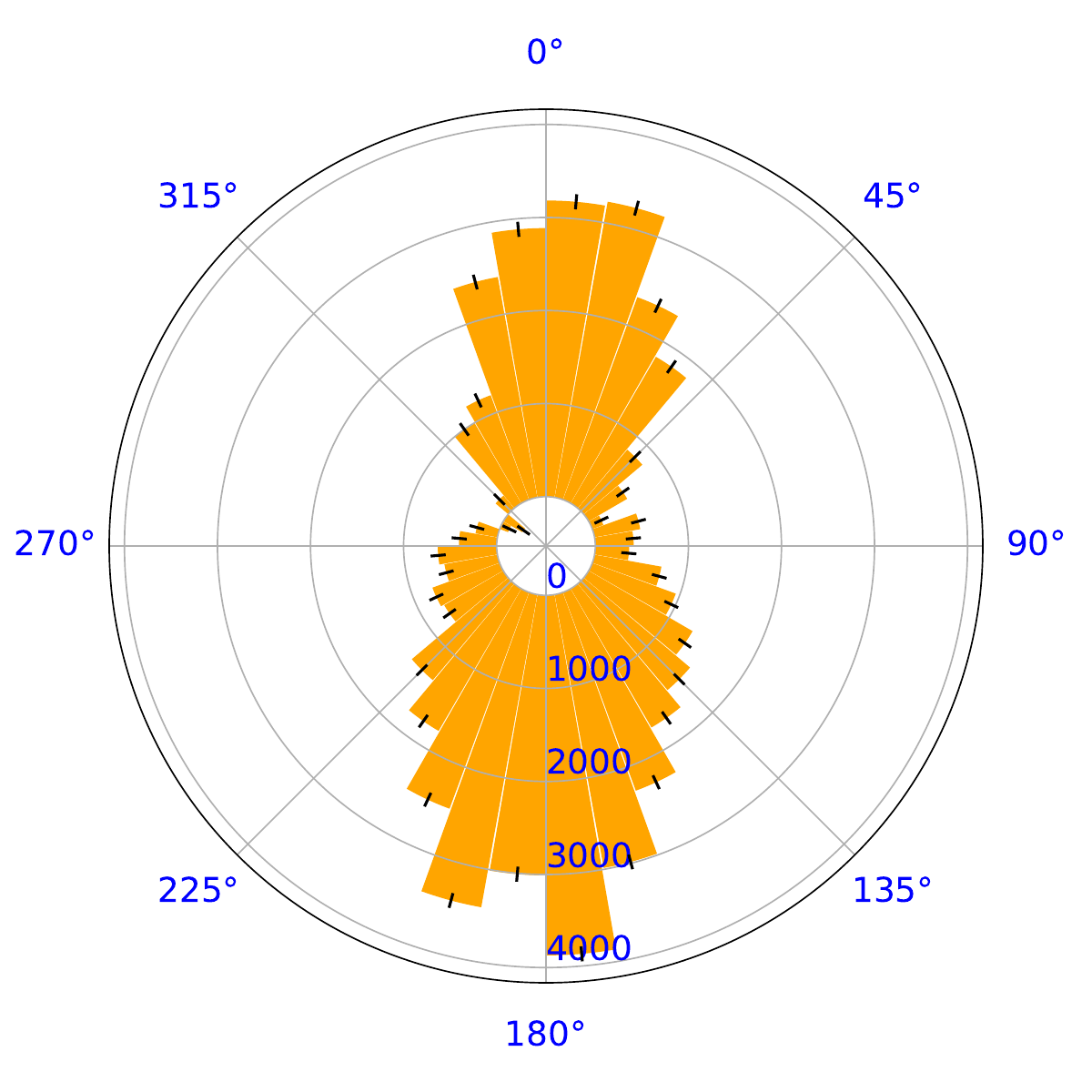}
    \includegraphics[width=0.32\linewidth]{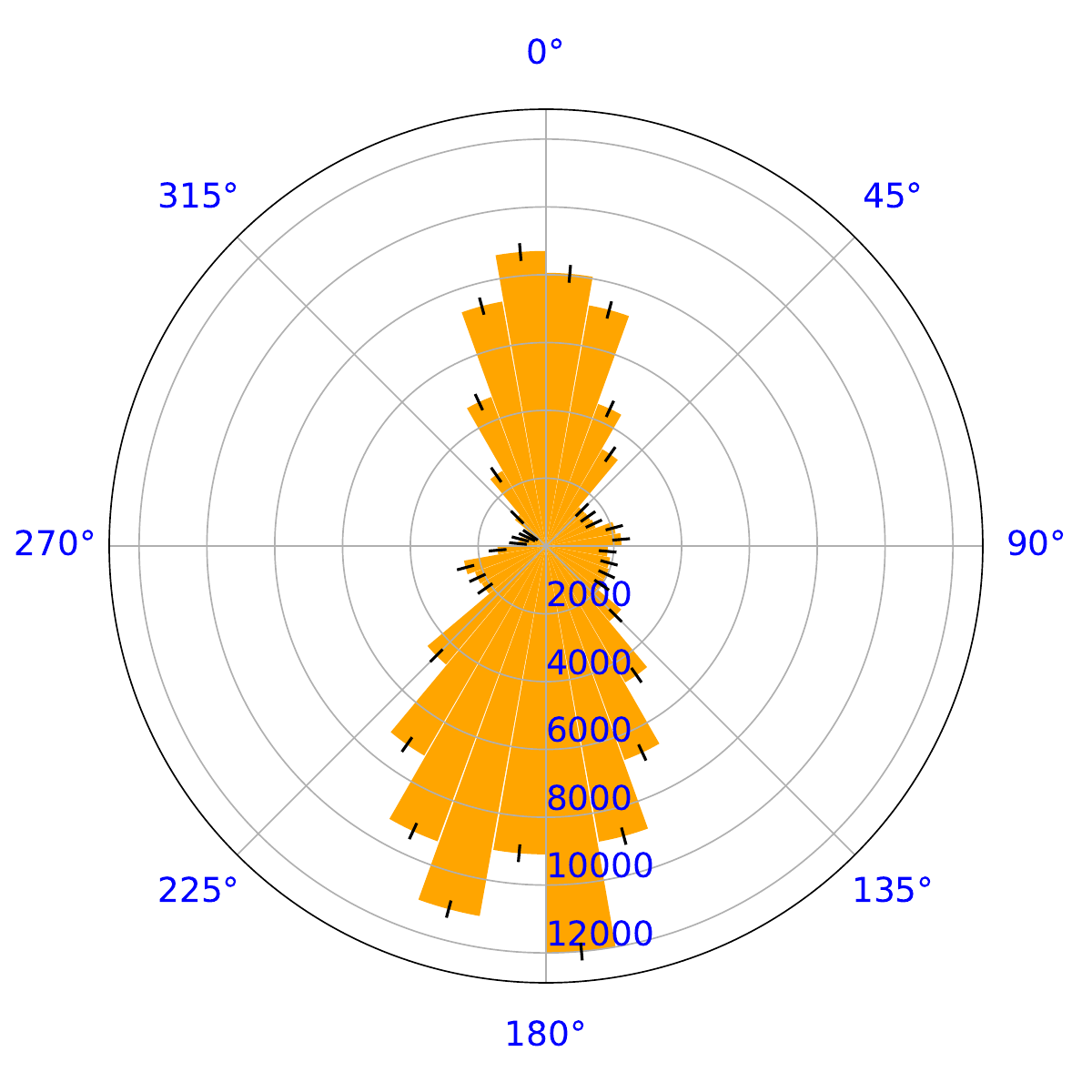} \\
    \caption{\textbf{Top:} Final KLIP-RDI reductions using MagAO+Clio data at $L$-band showing the inner disk component in a north-up, east-left orientation. The colorbar shows pixel values in counts/s. \textbf{Bottom:} Radar plots where the wedges portray the summed pixel counts in the above images in $10^{\circ}$ annular slices to quantify the  azimuthal flux in our KLIP-RDI images. The $36$ annular slices are defined in the region between the two dotted white lines, about $0\farcsec19$ to $0\farcsec57$, with the black ticks at the end of each orange wedge representing the $3 \sigma$ scatter on the pixel counts in each slice.}
    \label{fig:finalklip}
\end{figure*}

\subsection{Data Reduction and PSF Subtraction}
\label{sec:reduction}

To prepare our data for background subtraction, we divided each image by the number of coadds followed by a linearity correction using the procedure outlined in \cite{morzinski_magellan_2015}. We then divided each image by its exposure time to convert pixel values to counts/s. \textbf{To remove the sky and telescope thermal background in our images, we assumed that the sky emission was stable enough to use as suitable background estimates for the images at the other nod position.} For a given exposure, we created a background estimate using a median stack made from a sequence of counter-nod images that were captured closest in time. We employed a least-squares routine to determine the scale factor and offset that minimizes the residuals between this background estimate and a given target image ensuring that any stars or detector artifacts inconsistent between the two images were masked digitally. We extracted $128 \times 128$ pixel $(2" \times 2")$  images centered on the star and then used a phase cross correlation routine to register, with sub-pixel accuracy, all thumbnail images to the first exposure of \hd{} in the sequence \citep{guizar-sicairos_efficient_2008}.

Since MagAO provides exceptional PSF stability and delivers high Strehl ratios in the 1-5$\mu$m regime \citep{morzinski_magellan_2015}, we performed PSF subtraction using RDI coupled with the Karhunen-Lo$\grave{\text{e}}$ve Image Projection (KLIP; \citealt{soummer_detection_2012}) algorithm, which is an application of Principal Component Analysis that assumes discrete-to-discrete operations with Gaussian statistics. We made use of the Python-based \texttt{pyKLIP} package \citep{wang_pyklip_2015} to perform KLIP-RDI on our data with images of HD 144271 (see Section \ref{sec:obs}) serving as our reference PSF library.

The top row in Figure \ref{fig:finalklip} shows the final KLIP-reduced images for each passband. The bottom row shows radar plots illustrating the summed pixel values in annular slices between $0\farcsec19$ and $0\farcsec57$ as marked by the two white dotted lines in the images. These radar plots represent the azimuthal flux distribution detected in the inner disk region.

One of the challenges when using KLIP-RDI to remove starlight to reveal faint underlying circumstellar objects is choosing the optimal number of KLIP modes to use when forming the basis such that self- or over-subtraction is minimal \citep{milli_impact_2012}. \textbf{Given that this inner disk component has been directly imaged previously (\citealt{currie_matryoshka_2016}; \citealt{perrot_discovery_2016}; \citealt{mawet_characterization_2017}; \citealt{bruzzone_imaging_2020}; \citealt{singh_revealing_2021}) we optimized the $S/N$ of the brightest part of the previously known disk ansae by performing a grid search over the number of KLIP modes and the size of the central software star mask (i.e., the IWA).} We fixed the outer working angle (OWA) of the annular search region to OWA $ = 80 \text{ pixels} \approx 1\farcsec3$.

To estimate the noise maps that we used in calculating $S/N$ as well as in our forward model fitting, we followed \cite{lawson_scexaocharis_2020} and computed the noise at every radial location using concentric measurement annuli of 1 FWHM width to compute the standard deviation in each of our KLIP-RDI images. To improve our estimation, we made use of software masks to exclude the prominent north and south disk ansae in our noise calculation. We note that it is difficult to compute the underlying noise because there are few to no pixels at each radius that do not have disk flux. 

\begin{figure*}[h!]
    \centering
    \includegraphics[width=\linewidth]{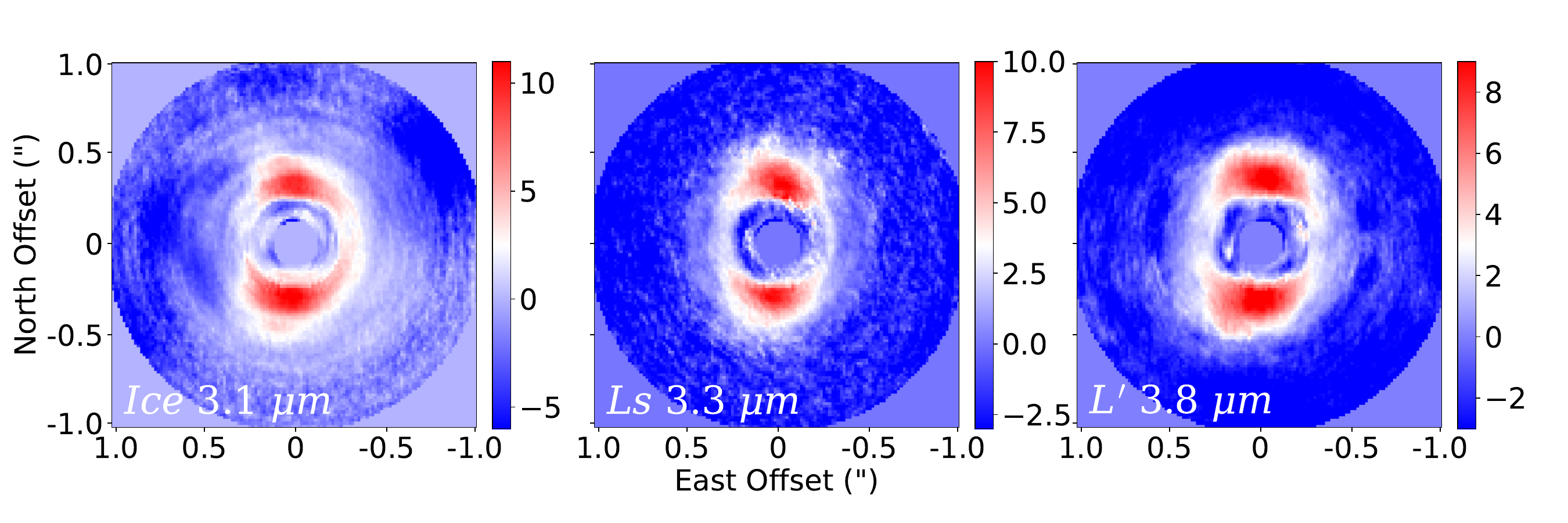}
    \caption{S/N per pixel maps of KLIP-RDI reduced images of the $L'$, $Ls$, and $Ice$ datasets where the central star region has been overlaid with a software mask of radius $r = 0\farcsec13$.}
    \label{fig:snre}
\end{figure*}

\begin{figure*}
    \centering
    \includegraphics[width=0.48\linewidth]{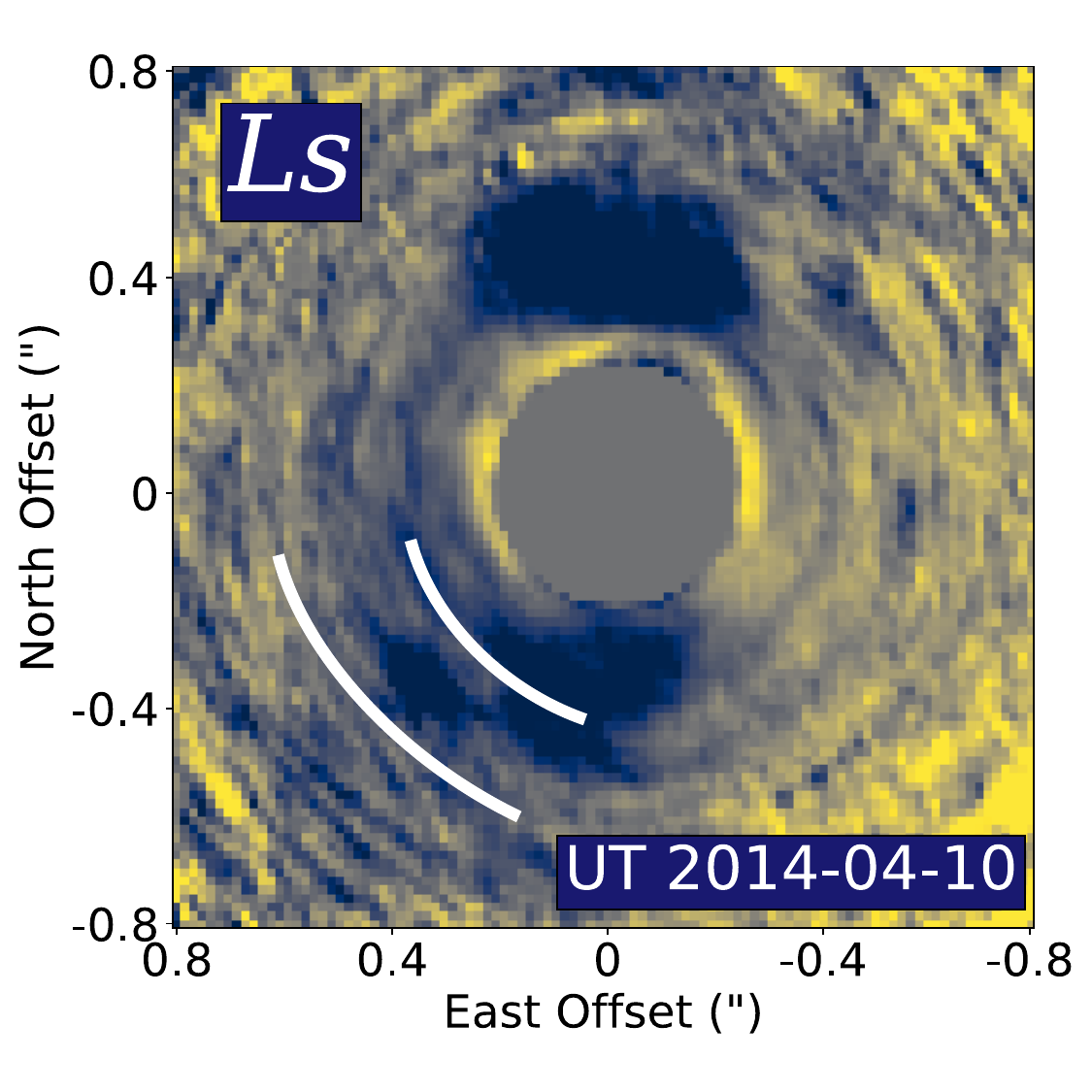}
    \includegraphics[width=0.48\linewidth]{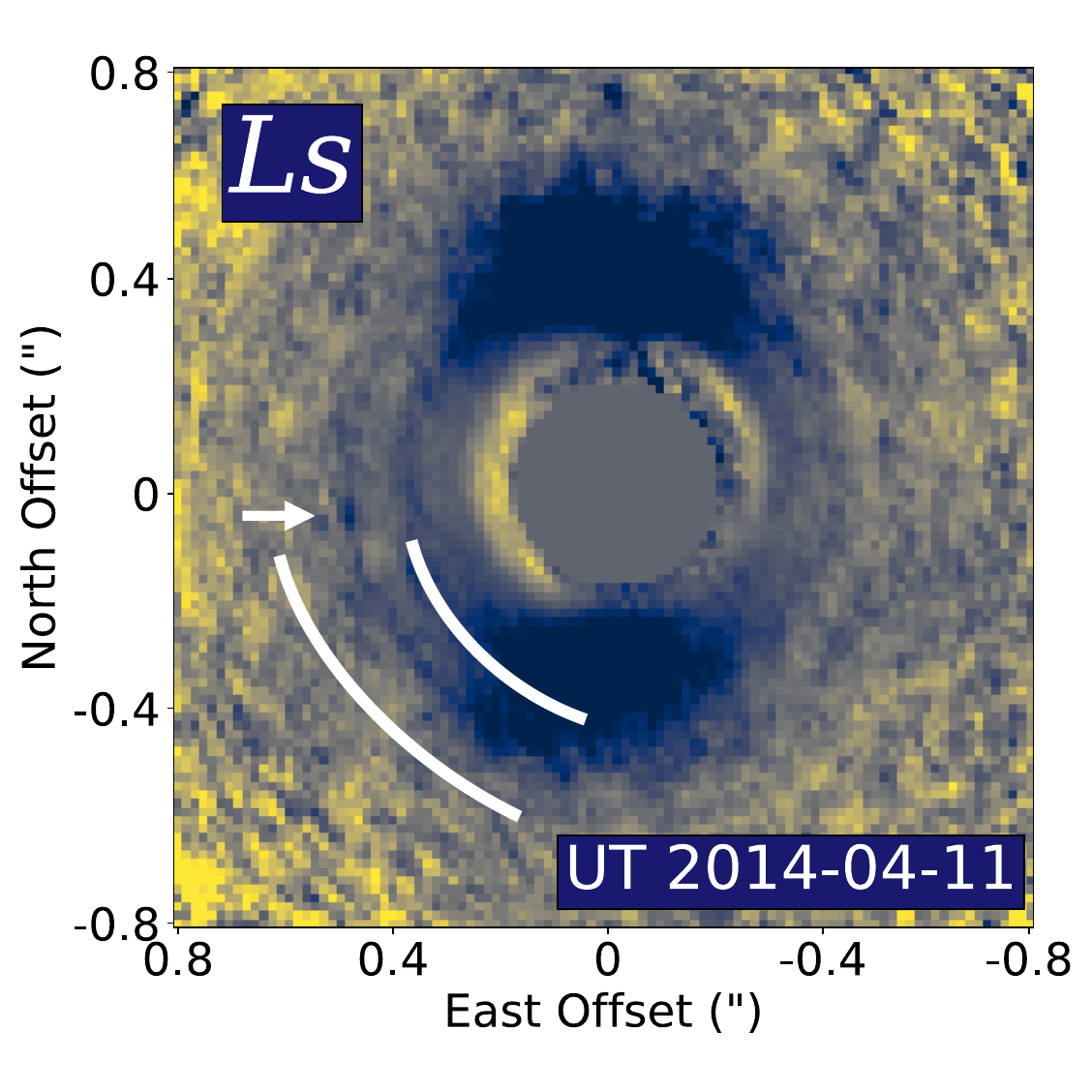}
    \caption{Inverted color map images of our observations at $Ls$ corresponding to the nights of UT 2014-04-10 to 11. The pixel values have been multiplied by $r^2$ to bring out faint structures in the outer regions of the disk. \textbf{White curves bound the arc or spiral-like structure to the southeast in each image, which was previously identified in \cite{mawet_characterization_2017}.} The right image, which corresponds to the higher quality dataset, hints at a point source-like feature at the terminus of the arc or spiral arm.}
    \label{fig:ls-feat}
\end{figure*}

The best S/N was achieved using an IWA of 8 pixels ($0\farcsec13$) coupled with a basis consisting of 3, 6, and 7 modes for $Ice$, $Ls$, and $L'$ datasets respectively. Figure \ref{fig:snre} shows the corresponding S/N results for our final reduced images. The full scope of this grid search is summarized in Table \ref{tab:tnr}. We note that for our $Ls$ datasets imaged on UT 2014 April 10 and 11 (start of night), we additionally optimized the KLIP reduction such that $S/N$ of a spiral/arc feature is maximized; these results are shown in Figure {\ref{fig:ls-feat}}. \textbf{Measuring the $S/N$ of the spiral/arc feature in the right panel image in Figure \ref{fig:ls-feat} from the white arrow down to the southern disk region yields around 1.5 to 2. We repeated the KLIP data reduction for this higher quality $Ls$ dataset from UT April 11 but instead used angular differential imaging to better preserve the footprint of a potential point source (marked in the right panel of Figure \ref{fig:ls-feat} with a white arrow). This analysis yielded a $S/N$ of 2.7 for the candidate point source using $27/60$ KLIP modes for the PSF subtraction.}

\begin{table}
\centering
\caption{Chosen KLIP Parameters Based on $S/N$ Optimization}
\label{tab:tnr}
\begin{tabular}{clccc}
\hline \hline
      & Date Obs. (UT)          & $N_{\text{modes}}$ & IWA $(")$ & OWA $(")$ \\ \hline
$L'$  & 2014 Apr 09             & 7/80              & 0.13      & 1.3       \\
$Ls$  & 2014 Apr 11 & 6/40               & 0.13      & 1.3       \\
$^{\dagger}Ls$  & 2014 Apr 10             & 8/20               & 0.22      & 1.3       \\
$^{\dagger}Ls$  & 2014 Apr 11 & 9/40               & 0.19      & 1.3       \\
$Ice$ & 2015 May 29             & 3/81              & 0.13      & 1.3       \\ \hline
\end{tabular}
\footnotesize
\raggedright

\vspace{1mm}
\textbf{Notes.} $N_{\text{modes}}$: number of KLIP modes; IWA: inner working angle; OWA: outer working angle. \\
$^{\dagger}$ KLIP parameters used to optimize the $S/N$ of spiral arm feature for Figure \ref{fig:ls-feat}.
\end{table}

\section{Disk Modeling}
\label{sec:modeling}

\subsection{Model Description}
\label{sec:model_desc}

We performed forward modeling to further characterize the geometry of the inner disk using our total intensity images taken on the nights of UT 2014-04-09 $(L')$, UT 2014-04-11 $(Ls)$, and UT 2015-05-29 $(Ice)$ (see Section \ref{sec:obs}).


Following \cite{chen_multiband_2020}, we explored the geometrical parameters and grain sizes in the 20 to 70 au region of the inner disk by implementing the DiskFM pipeline \citep{mazoyer_diskfm_2020} of the \lib{pyKLIP} package which accomplishes forward modeling based on the principles presented in \cite{pueyo_detection_2016}. Additionally, we used the Markov-Chain Monte Carlo (MCMC) sampler package, \pkg{emcee} \citep{foreman-mackey_emcee_2013}, as a wrapper to DiskFM for estimating the parameter values.

\textbf{We generated scattered light disk models at $3.8 \mu$m, $3.3 \mu$m, and $3.1 \mu$m to use in our forward modeling and assumed that the inner disk is optically thin (\citealt{bruzzone_imaging_2020}; \citealt{thi_gas_2014}) to interpret model fits to the observed brightness asymmetry.} Our final models were limited to a single ring geometry since preliminary trials with multi-ring models (inspired by the ringlets presented in \citealt{perrot_discovery_2016}) produced poor fits to our data. We made use of the disk model in stellocentric coordinates from \cite{ren_exo-kuiper_2019}, which is a modified form of that in \cite{millar-blanchaer__2015}:

\begin{equation}
    \rho(r) \propto \left[\left(\frac{r}{r_c}\right)^{-2\alpha_{\text{in}}} + \left(\frac{r}{r_c} \right)^{-2\alpha_{\text{out}}} \right]^{-\frac{1}{2}},
\end{equation}

where $r$ is the radial coordinate, $r_c$ is the critical radius of the disk corresponding to the location of the peak dust density, and $\alphain$ and $\alphaout$ denote the radial power laws of the surface brightness interior and exterior to the critical radius, respectively. Note that we constrained $\alphain$ to be positive and $\alphaout$ to be negative for our modeling. We modeled the vertical distribution $Z(r,z)$ (normal to the disk midplane) using a Gaussian profile, i.e., 

\begin{equation}
    Z(r,z) \propto \exp{\left[-\left(\frac{z}{h_0r^{\beta}} \right)^2 \right]},
\end{equation}

where $z$ is the vertical axis coordinate, $h_0$ is the constant aspect ratio, and $\beta$ is the dust flaring component. We arbitrarily fixed the disk aspect ratio to $h_0 = 0.02$ and the dust flaring exponent was set to $\beta = 1$ (non-flaring), \textbf{supported by our above-mentioned assumption that the dust is optically thin.} 
As also found by \cite{mawet_characterization_2017}, our data allow limited constraint on the dust density radial power law indices $\alphain$ and $\alphaout$, so we adopted $\alpha_{in} = 1.5$ and $\alpha_{out} = -9$ by visual inspection of the forward models.

\begin{table*}
\centering
\caption{DiskFM Modeling Results for \hd{}}
\hspace*{-2cm}\begin{tabular}{lccccccc}
\hline \hline
\multirow{2}{*}{Parameter$^\text{a}$} &  & \multirow{2}{*}{Range Explored$^\text{b}$} &  & \multicolumn{3}{c}{Best Fit Values$^\text{c}$}   & \multirow{2}{*}{\cite{mawet_characterization_2017}}                                                                    \\ \cline{5-7} 
                           &  &                                 &  & $Ice$                             & $Ls$                              & $L'$   &                           \\ \hline
$\log_{10}R_c$ (au)                 &  & (25, 50)                          &  & $40.39 \pm 0.50$ & $39.62 \pm 0.43$ & $39.07 \pm 0.53$ & $39 \pm 4$ \\
$i (^{\circ})$             &  & (30, 80)                          &  & $47.47 \pm 1.32$ & $47.61 \substack{+0.98 \\ -1.04}$ & $48.10 \substack{+0.97 \\ -1.07}$ & $53 \pm 6$ \\
$PA (^{\circ})$            &  & (-20, 20)                         &  & $-2.45 \substack{+0.90 \\ -0.93}$ & $-2.19 \pm 0.90$                   & $-0.68 \substack{+0.92 \\ -0.94}$ & $-11 \pm 8$ \\
$dx$ (au)                  &  & (-15, 15)                         &  & $-0.58 \pm 0.57$ & $-5.99 \substack{+0.78 \\ -0.74}$ & $-3.14 \substack{+1.30 \\ -1.50}$ & $-2 \pm 7$ \\
$dy$ (au)                  &  & (-10, 10)                         &  & $1.23 \substack{+0.55 \\ -0.53}$ & $1.96 \pm 0.41 $ & $2.34 \substack{+0.46 \\ -0.48}$                  & $0 \pm 4$\\
$\log_{10}a_{\text{min}}$ ($\mu$m)   &  & (0.001, 10)                       &  & $0.62 \substack{+0.02 \\ -0.01}$                   & $0.66 \substack{+0.02 \\ -0.01}$  & $0.79 \substack{+0.03 \\ -0.04}$                 & $0.1$  \\ \hline
$F$ (mJy)$^\text{d}$ & & & &  $19.48 \pm 0.22$ & $26.34 \pm 0.23$ & $31.31 \pm 0.27$ & \\ \hline
\end{tabular}

\footnotesize
\raggedright
\vspace{1mm}
\textbf{Notes.}  \\
\vspace{1mm}
$^{\text{a}}$ We included parameter $N$ in our modeling to handle the flux normalization, but it is not shown in this table. \\
$^{\text{b}}$ We assumed uniform prior distributions for all parameters except the disk midplane offset $dx$ which required Gaussian prior regularization due to parameter space degeneracies. \\
$^\text{c}$ 16th, 50th, and 84th percentiles. \\
$^\text{d}$ We include the total flux of the modeled disk in mJy units which we computed by summing the best-fit model; see text. \\
\label{tab:diskfm}
\end{table*}

We generated the intensity of the model by computing a brightness integral for each pixel $(x',y')$ along the line of sight $(z')$ (where primed coordinates denote the detector frame) expressed as

\begin{equation}
    I(x',y') = I_0 + \int_{z' = -R_2}^{R_2} dz' \frac{N_0}{r^2}  \rho(r) Z(r,z) P(\theta),
\end{equation}

where $P(\theta)$ is the scattering phase function (SPF) and $R_2$ marks the user-set outer boundary where the dust density defaults to zero \textbf{(i.e., the observed outer radius of the disk). To speed calculations, we defined $R_2$ to correlate tightly to the edge of the observed disk signal in our KLIP-RDI images since we assume the radial extent of the disk along the major axis is the largest spatial dimension.} This resulted in an elliptical modeling region with a major axis of $0\farcsec54$ and minor axis of $0\farcsec34$ and rotated such that the major axis aligns with north-south; $R_2$ corresponds to the outer boundary for the minimization region as shown as the dashed white ellipse in the KLIP-RDI images in Figure \ref{fig:master_diskfm_results}. Additionally, $I_0$\footnote{Preliminary modeling confirmed that $I_0 = 0$ since we removed the sky background prior to performing the KLIP-RDI procedure. So we fixed $I_0 = 0$ for our final modeling results.} and $N_0$ are parameters representing a constant offset and scale factor to handle flux normalization respectively. The disk coordinate system undergoes tilting with respect to the observer by varying an inclination parameter $i$ and the position angle $PA$ manages the clocking with respect to the sky. The scattering angle $\theta(x',y',z')$ is a function of position in the image.

We used a custom SPF, which is calculated from simulated agglomerated debris particles (ADPs) (see \citealt{zubko_effect_2015}; \citealt{arnold_stumbling_2022} for a full description). The irregular, porous shapes of these ADPs (porosity of 70-80\%) attain a more realistic representation of the shapes of dust grains than a distribution of, e.g., solid or hollow spheres. We created a dust grain size distribution $n_a$ based on a minimum grain size $\amin$, a maximum grain size $\amax$, and a power law index $q$ such that $dN/da \propto a^{-q}$. We then obtain a complex index of refraction, typical of astronomical silicates, using results from \cite{draine_scattering_2003} and \cite{draine_scattering_2003-1} by linearly interpolating at our wavelength of interest. Since the lookup tables from \cite{arnold_stumbling_2022} parameterize the scattering efficiencies  by complex index of refraction and size parameter, we use the results from the above steps to query the scattering efficiencies for angles $\theta = [0,\pi]$ to obtain the custom SPF. For our modeling, we fixed the maximum grain size to $\amax = 1$ cm and the size distribution exponent to $q = 3.5$, but assigned the minimum grain size $\amin$ as a free parameter.

In total, we fit for 7 parameters in our forward modeling efforts: the critical radius $R_c$, the inclination $i$, the position angle $PA$, the stellar offsets in the disk's midplane $xy$-directions $dx$ and $dy$, the flux normalization $N$, and the minimum grain size $a_{\text{min}}$.

Importantly, we note that we chose a simple scattered-light disk model appropriate for measuring the KLIP-RDI throughput and observed disk surface brightness. We expect that our model may not match more detailed physical attributes of the disk.  

\subsection{MCMC Parameter Estimation}
\label{sec:mcmc}

To find the key geometric properties that best fit our reduced disk images, we used the Markov-Chain Monte Carlo (MCMC) sampler \texttt{emcee} package \citep{foreman-mackey_emcee_2013}. The range of values probed for all fitted parameters are shown in Table \ref{tab:diskfm} with the exception of the model flux scaling parameter $N$. In short, our simple disk models are forward modeled using DiskFM \citep{mazoyer_diskfm_2020} to estimate the algorithmic artifacts and throughput loss using KLIP. An overview of the full forward modeling procedure is as follows:

\begin{enumerate}
    \item Generate a disk model using the method outlined in \ref{sec:model_desc}.
    \item Convolve the disk model with the empirically-measured, unsaturated instrument PSF (see Sec. \ref{sec:obs}) to simulate a model image.
    \item Use the \texttt{pyKLIP} DiskFM pipeline to produce a forward model (FM) using the same KLIP parameters as the final KLIP-RDI image.
    \item Gauge the goodness of fit of these results using the standard $\chi^2$ metric:
    \begin{equation}
        \chi^2 = \sum_S \frac{(\text{Data} - \text{FM})^2}{\text{Uncertainty}^2},
    \end{equation}
\end{enumerate}

where $S$ is the zone of interest where $e^{-\chi^2/2}$ is maximized; see Section \ref{sec:results_modeling}. Our estimate for the uncertainty is described in Section \ref{sec:reduction}.

To assure convergence, we repeated the enumerated steps above until the number of iterations exceeded 50 times the autocorrelation time of the chains. This amounted to modeling 564.0k, 808.4k, and 611.0k disk models for $L'$, $Ls$, and $Ice$ bands respectively. We assumed uniform prior probability distributions for all fitted parameters except $dx$, which we regularized with a Gaussian prior ($\sigma = 2$ au) centered on zero. Without regularization, degeneracies in the parameter space resulted in disk models we regarded as non-physical. Additionally, in a burn-in phase, we ignore the number of initial steps in each chain equal to twice the autocorrelation time, as recommended in the documentation for \texttt{emcee}\footnote{\url{https://emcee.readthedocs.io/en/stable/user/autocorr/}}. 

\subsection{Photometric Calibration}
\label{sec:phot_calib}
We use our short, unsaturated images of HD 141569 for photometric calibration. We performed sky background subtraction on these images using the same procedures as in Section \ref{sec:reduction}, divided by the exposure time, and then summed the counts in an aperture of radius $r = 3\farcsec2 (\approx 30 \lambda/D$). We inspected curves of growth \textbf{(i.e., summed counts as a function of aperture radius)} for each background-subtracted calibration frame and excluded results from frames where the curve did not asymptote, which indicated that the background subtraction was sub-optimal. We then averaged our aperture photometry for non-rejected calibration frames.

\begin{figure}
    \centering
    \includegraphics[width=\linewidth]{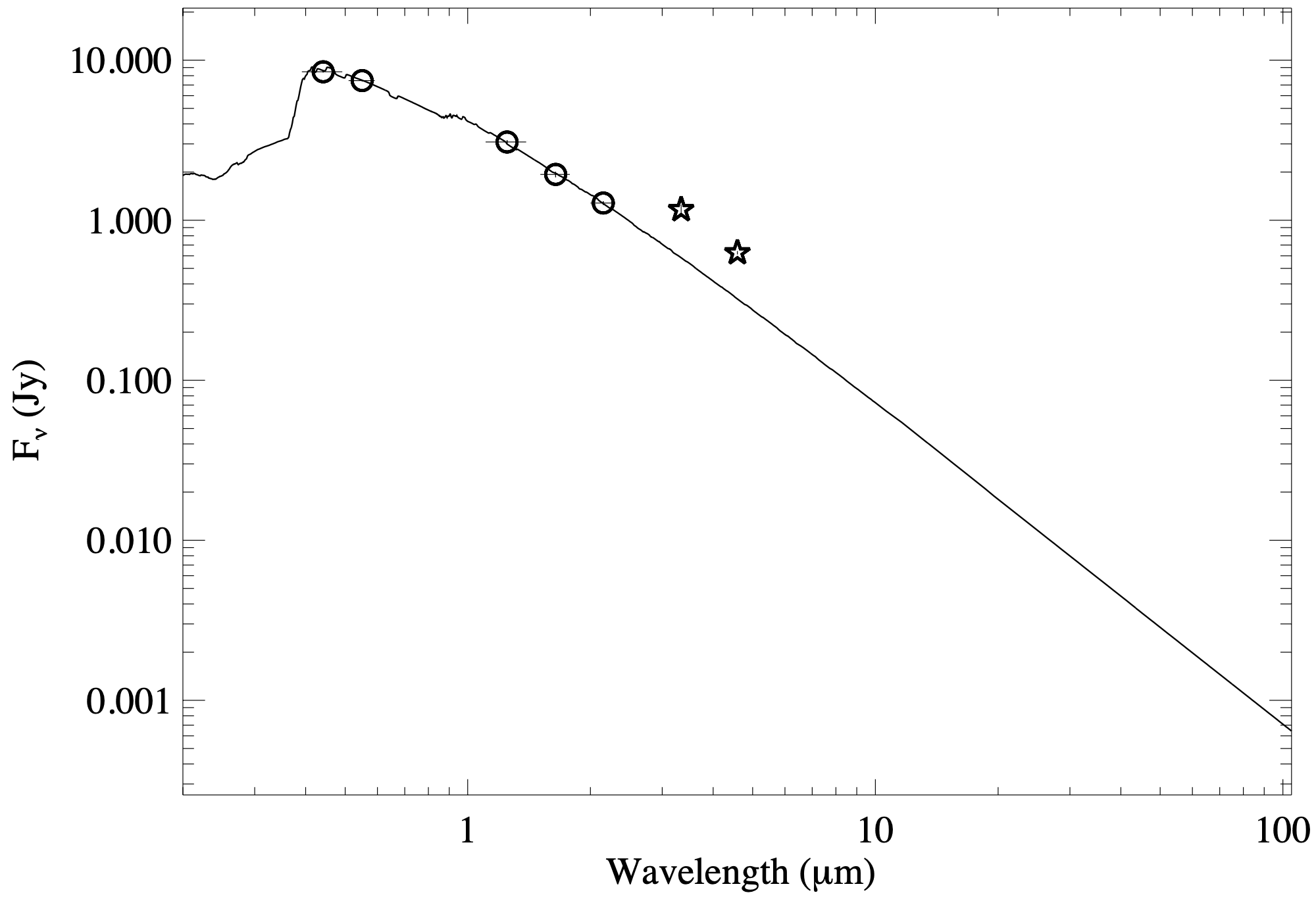}
    \caption{Kurucz model fit ($T_{\text{eff}} = 10000$ K, $\log g = 4.5$); with a reddening extinction correction of $A_V = 0.35$ mag) to \hd{} BVJHK catalog magnitudes which are represented by the circles. We plot AllWISE \citep{cutri_vizier_2021} W1 and W2 photometry as stars to show a poor fit to our model compared to the visible and near-infrared data. We ignore the AllWISE photometry when calculating our photometric calibration.}
    \label{fig:sed}
\end{figure}

\begin{table}[htbp]
\centering
\caption{Comparison of photometric calibration factors calculated using images of HD 141569 A and PSF reference HD~144271 to convert pixel values from counts/s to Jy/arcsec$^2$.}
\label{tab:conversion}
\begin{tabular}{lccc}
\hline
\hline
                     & $Ice$             & $Ls$              & $L'$              \\ \hline
HD 141569 A & 2.57e-11 & 7.19e-12 & 2.43e-12 \\
HD 144271            & 2.59e-11          & 7.33e-12          & 2.54e-12          \\ \hline
\end{tabular}
\end{table}

\textbf{To convert our final reduced images and disk models from counts/s to mJy/arcsec$^2$, we use the ratio of the model flux density (in Jy units) to the average counts/s measured in our unsaturated calibration exposures of HD 141569A. To calculate HD~141569A's flux density in each of our filters, we model its photosphere with a Kurucz model ($T_{\text{eff}}, \log g = 4.5$) and fitted it to catalog $BV$ (Tycho-2; \citealt{hog_tycho-2_2000}) and $JHK$ (2MASS; \citealt{skrutskie_two_2006}) magnitudes of \hd{} with a reddening extinction correction of $A_V = 0.35$ mag. We note that we ignore WISE W1 and W2 photometry because they constitute a poor fit to our model possibly due to contamination from the two M-dwarf companions as shown in Figure \ref{fig:sed}. As a sanity check and to ensure that excess infrared flux from the disk was not significantly contaminating our photometry, we repeated the above methods for calibration frames of the reference star HD 144271 and obtained calibration factors to within about 1\%, 2\%, and 4\% for $Ice$, $Ls$, and $L'$ filters respectively. All calibration factors are shown in Table \ref{tab:conversion} for comparison.}



\section{Results} 
\label{sec:results}

\begin{figure*}
    \centering    \includegraphics[width=0.99\linewidth]{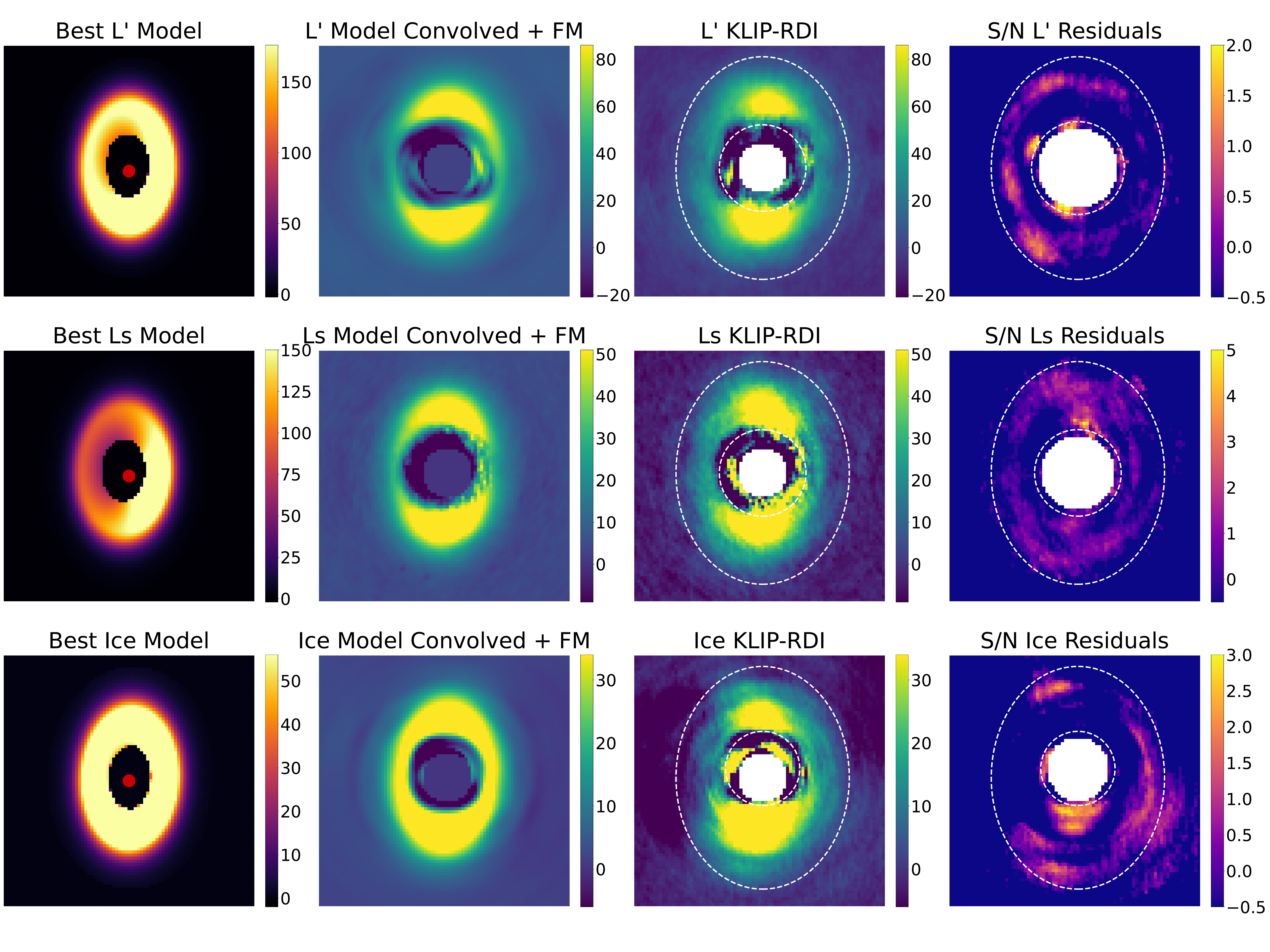}
    \caption{Forward modeling results for $L'$, $Ls$, and $Ice$ images where the color bar for the three leftmost columns are in surface brightness units of mJy/(")$^2$. The red circle in the best model images in the first (leftmost) column indicate the location of the star. We mark the zone where $\chi^2$ is evaluated between the dashed white lines in the KLIP-RDI images shown in the third column. The we used the best-fit models in the leftmost column to compute the flux density of the disk by multiplying the models by the squared pixel scale and summing the resulting pixel values; we appended the resulting values to Table \ref{tab:diskfm}. \textbf{The best-fit models for the $Ice$ and $L'$ bands assume nearly isotropic scattering while the model for the $Ls$ passband is highly asymmetric due to a stellar offset in the x-direction. The front side of the disk is the eastern (left) half.}}
    \label{fig:master_diskfm_results}
\end{figure*}

\subsection{Potential Spiral Arm and Point-like Feature}
\label{sec:results_spiral}

Notable features in our $Ls$ KLIP-RDI images include an arc or spiral-like feature to the southeast, similar to what was seen by \cite{mawet_characterization_2017} and \cite{currie_matryoshka_2016} at $L'$, by \cite{perrot_discovery_2016} at shorter wavelengths in the near-infrared, and by \cite{konishi_discovery_2016} in the visible. Figure \ref{fig:ls-feat} shows our $Ls$ KLIP-RDI images that were reduced with specific KLIP parameters to bring out faint spiral or arc feature (see Section \ref{sec:reduction}) and where each pixel was multiplied by $r^2$ to better show the dust density distribution and faint outer features. The arc or spiral feature is traced by red arrows. The image from observations on UT 2014-04-11 clearly show an arc or spiral feature originating in the southern-most region of the inner disk and trailing up to the northeast, terminating at $\sim0\farcsec5$ to the east of the star. Interestingly, there appears to be a point-like source at the end of the arc or spiral feature. Although the night of UT 2014-04-10 presented poor observing conditions resulting in a final KLIP-RDI image of poor quality, the arc or spiral-like feature is still discernible but the point-like source is not. These features are not recovered at any other passbands in our data.

\begin{figure*}
\centering
\begin{tabular}{ccc}
         \includegraphics[width=\linewidth]{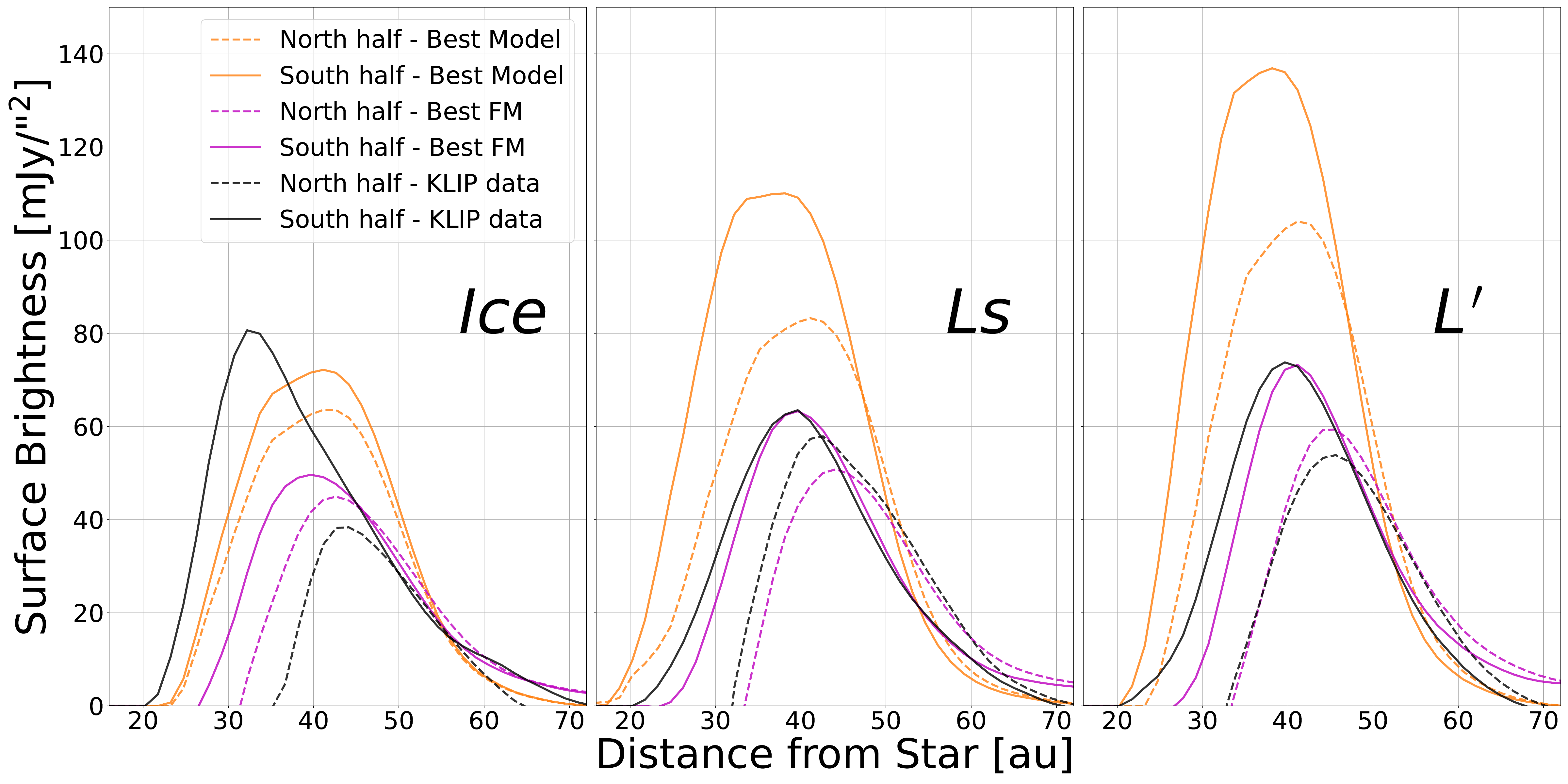}
    \end{tabular}
    \footnotesize \caption{Surface brightness profiles extracted along the major axis using aperture photometry ($r_{\text{ap}} = \frac{\lambda}{2D} \approx 0\farcsec06$) of the best-fit models for our \hd{} disk images at $Ice$ (3.1 $\mu$m), $Ls$ (3.3 $\mu$m), and $L'$ (3.8 $\mu$m). Dashed lines represent profiles across the north half of the disk while solid lines represent profiles for the south half. We denote profiles for the best-fitting disk model, best-fitting model convolved then forward modeled, and the final KLIP-RDI reduced image with orange, magenta, and black lines respectively. \textbf{The best-fitting model profiles represent the disk photometry without the throughput loss due to the KLIP algorithm.}}
    \label{fig:sb-profiles}
\end{figure*}

\subsection{Modeling results}
\label{sec:results_modeling}

The final corner plots from our analysis as well as further information on the modeling parameters are shown in Figure \ref{fig:corner_38}, Figure \ref{fig:corner_33}, and Figure \ref{fig:corner_31} in Appendix \ref{sec:appendixA}. For each passband, Figure \ref{fig:master_diskfm_results} shows our best-fit disk models, their forward models, the final KLIP-RDI images, and the S/N of the residuals in columns one through four respectively. We mark the location of the star with a red circle and bounded the zone where $\chi^2$ was minimized with white dashed lines. Note that some signal appears to be excluded interior to the inner boundary of this zone, but the extent to which the non-negative pixel values in these high speckle noise regions represent the inner disk is unclear.

\subsubsection{Disk Geometry}
\label{sec:results_geo}

The best-fit parameters from our forward modeling are summarized in Table \ref{tab:diskfm} along with a comparison to $L'$ results from \cite{mawet_characterization_2017}. Best-fit values for the geometrical parameters are consistent within $3\sigma$ between the 3 passbands with the exception of the $dx$ stellar offset. \textbf{We assume that the east side of the disk is closest to us from polarimetric imaging results (\citealt{bruzzone_imaging_2020}; \citealt{singh_revealing_2021}) and thus should exhibit strong forward scattering (assuming realistic dust grain properties) relative to the west side.} However, our best fit models at $L'$ and $Ls$ exhibit a lower flux on the eastern half. The model accomplishes this by inducing $dx$ and $dy$ stellar offsets (i.e., the disk is shifted along its midplane) thus placing the northeast part of the disk farther from the star.  The $L'$ and $Ls$ models hint at an offset of about 3 and 6 au in the negative $x$-direction in our stellocentric coordinates while the $Ice$ model prefers an offset  of about 0.5 au in the negative $x$-direction, which is significantly different from the $Ls$ result. \textbf{These offsets could manifest in the model fit due to other physical phenomena; see below (Section \ref{sec:discuss-morph})}. However, this modest eastward offset for the $Ice$ model may be due to the circular oversubtraction artifact (see Figure \ref{fig:finalklip}). Despite these disagreements in best-fitting $dx$ values, they are all consistent with the $L's$ model from \cite{mawet_characterization_2017}. We discuss implications of asymmetric dust density or gas in the disk from these stellar offsets below in Section \ref{sec:discuss-morph}. 

S/N maps of the residuals between the best forward model and the KLIP-reduced data of the inner disk show an enhancement in the $Ice$ passband in the south ansa region of the disk that peaks at $\sim3\sigma$; this same feature appears at $L'$ in the residuals in \cite{mawet_characterization_2017}. The residuals map at $Ls$ also displays a curiously small feature at $\sim5\sigma$ significance to the north and immediately outside the central star mask. However, this may be a false positive given the small spatial extent of this feature and that it resides close to the high speckle noise region close in to the star. Similar features are seen in the $L'$ residuals map at roughly the same radial distance, but at a lower significance.

\subsubsection{Dust Grain Properties}
\label{sec:results_grains}

For all our models, we fixed the maximum grain size to $\amax = 1$ cm and the size distribution exponent to $q = 3.5$. We obtained a best-fit minimum grain size $\amin$ of $0.62 \substack{+0.02 \\ -0.01} \mu$m, $0.66 \substack{+0.02 \\ -0.01} \mu$m, and $0.79 \substack{+0.03 \\ -0.04} \mu$m for $Ice$, $Ls$, and $L'$ bands respectively suggesting a majority of the scattering is due to sub-micron grains. Though the median values of these best-fit values differ between the 3 passbands, they are all consistent to within $3\sigma$. Further, this specific combination of grain parameters results in a disk model that scatters more or less isotropically which is perhaps unsurprising since our data probes a limited range of scattering angles (i.e., we only measure high S/N near the disk ansae). For completeness, we also performed forward modeling using a Henyey-Greenstein SPF which yielded a best-fitting disk model that isotropically scatters, but we omit those results from this paper in favor of results from the more realistic SPF calculated using the ADP scattering efficiencies from \cite{arnold_stumbling_2022}. Nonetheless, these results are consistent with the analysis shown in \cite{currie_matryoshka_2016} and \cite{mawet_characterization_2017} who also found isotropically-scattering disk models to best fit their total intensity images at $L'$.

\subsubsection{Disk Photometry}
\label{sec:results_sb}

We extracted the disk surface brightness along the major axis in our best-fit models and KLIP-RDI reduced images using aperture photometry with aperture size set to $r = \lambda/(2D) \approx 0\farcsec06$. The resulting profile curves are shown for the best-fit models (orange lines), forward models (magenta lines), and the reduced images (black lines) in Figure \ref{fig:sb-profiles}. We note that the degree of North-South brightness asymmetry is clearly portrayed among the three groups of profiles. Specifically, the $Ice$ band KLIP-RDI profiles show the greatest North-South asymmetry with the south half of the disk more than twice as bright as the north side (along the major axis), which our disk model is unable to reproduce (as noted in Section \ref{sec:results_geo}).

\begin{figure}
    \centering
    \includegraphics[width=\linewidth]{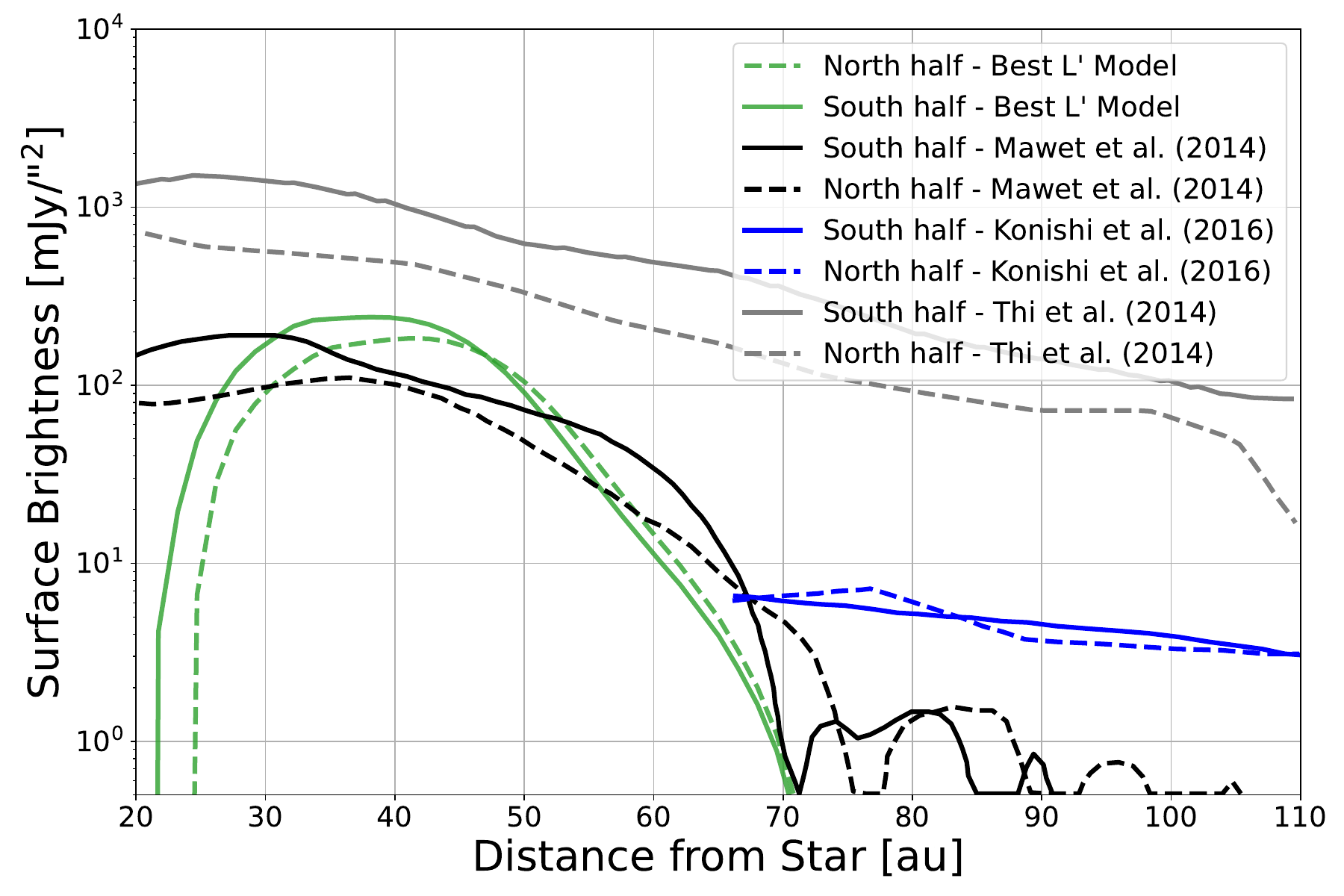}
    \caption{Comparison of our surface brightness measurement (in mJy/(")$^2$) with literature values in the region around the star between 20 au to 110 au. The dashed green and solid green curves correspond to the north and south region profiles of our best-fit disk model, respectively, measured along the disk major axis using aperture photometry with aperture radii set to $r = \lambda/(2D) \approx 0\farcsec06$. \textbf{The black curves show the published $L'$ surface brightness also measured using aperture photometry along the major axis from \cite{mawet_characterization_2017}. The blue curves show the surface brightness profile measured using broad-band optical imaging from \cite{konishi_discovery_2016} and the gray curves illustrate the $8.6 \mu$m PAH emission profile from \cite{thi_gas_2014}.}}
    \label{fig:sb_lit_comparison}
\end{figure}

Figure \ref{fig:sb_lit_comparison} compares the surface brightness profile extracted from our best-fit $L'$ model (green lines) with other profiles found in the literature. MagAO data show good agreement with the $L'$ profile extracted by \cite{mawet_characterization_2017}. 

\begin{figure}
    \centering
    \includegraphics[width=\linewidth]{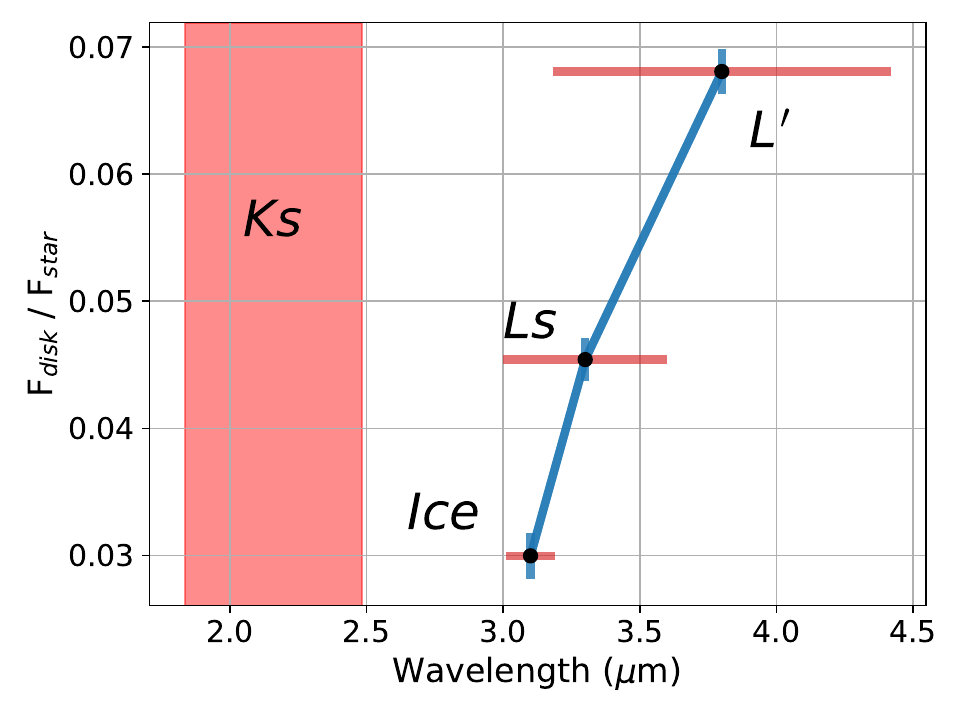}
    \caption{Disk spectrum with respect to the star using our best-fit $L'$, $Ls$, and $Ice$ disk models to compute the flux density (see text). The black points denote the disk flux density normalized by the estimated flux density of the star with the vertical blue bars showing the error. The horizontal red lines illustrate the filter bandwidths. We did not detect the disk at $Ks$, but we include the filter bandwidth on the left side of the plot. The flux density decrease from $L'$ to $Ice$ bands could indicate absorption of scattered light due to water ice at $3.1 \mu$m.}
    \label{fig:disk_spec}
\end{figure}

Using our best-fit disk models, we computed the total integrated flux in $F_{\nu}$ units of mJy which we append to Table \ref{tab:diskfm} (see Section \ref{sec:phot_calib}). These flux densities normalized by the estimated flux densities of the star at their respective passbands are shown in Figure \ref{fig:disk_spec}. The dip in brightness from $L'$ to $Ice$ mentioned above is clearly depicted.

\section{Discussion} 
\label{sec:discussion}

\subsection{Disk Morphology}
\label{sec:discuss-morph}

Our KLIP-RDI images reveal a significant north-south brightness asymmetry in the inner disk, which has also been previously identified (\citealt{singh_revealing_2021}; \citealt{bruzzone_imaging_2020};  \citealt{mawet_characterization_2017}; \citealt{perrot_discovery_2016}); the south half of the disk is brighter. Our model attempts to accommodate this by modifying the sizes of the dust grains to change the amount of forward versus backward scattering of the light and/or altering the irradiance at different regions by applying stellar offsets. We are able to reproduce the observed asymmetry at $L'$ and $Ls$ with midplane offsets such that the southwest region of the disk is closest to the star. An eccentric disk  could also cause an apparent offset. High eccentricity, (as seen in, e.g., HD 534143; \citealt{macgregor_alma_2022}) may result from the gravitational influence from nearby or embedded companions. Alternatively, having a higher density of scattering dust in the western half of the disk would make it brighter. The latter notion is consistent with results from \cite{singh_revealing_2021}, who modeled a higher dust density in the southwest region (roughly $220^{\circ}$ to $238^{\circ}$) to reproduce their polarized intensity images.

The aforementioned brightness asymmetry observed in our $Ice$-band image resulted in a wedged-shaped feature south of the central mask in the S/N of the residuals (bottom right panel of Figure \ref{fig:master_diskfm_results}; where the S/N peaks at $\sim3$). To quantify the residual flux, we created a bespoke aperture that was a good match to this feature to measure the residual flux density using the (data - model) image. This procedure yielded about $1.2$ mJy of additional flux density not in the forward model. However, we note that this area is in a region plagued by significant speckle noise in the $L'$ and $Ls$ KLIP-RDI images. While we attribute this excess flux to the disk, it is possible that it is just speckle noise.

Our fits to the disk inclination  are $\sim10^{\circ}$ lower than typical literature values (see, e.g., \citealt{currie_matryoshka_2016}; \citealt{perrot_discovery_2016}; \citealt{bruzzone_imaging_2020}; \citealt{singh_revealing_2021}) though consistent within uncertainties with the results from $L'$ image modeling by \cite{mawet_characterization_2017}. Our model fits may be driven to lower inclination through a combination of factors. First, the noise in the east and west regions close to the star may be underestimated because there are few pixels not containing disk flux with which to measure the noise (see Section \ref{sec:reduction}); the zone where $\chi^2$ is minimized includes these east-west regions. These minimization zones are illustrated in the third column in Figure \ref{fig:master_diskfm_results}. Second, as mentioned at the beginning of Section \ref{sec:modeling}, these discrepancies could be a consequence of using a simplistic disk model with some parameters that are not well constrained.

Our KLIP-RDI images at $Ls$ reveal an arc or spiral-like structure at the southeast boundary of the inner disk system; this has been previously noted in the $L'$ image studied in \citep{mawet_characterization_2017}. Though not explicitly identified, some semblance of this structure is also seen in visible band images \citep{konishi_discovery_2016}, and other near-infrared images \citep{perrot_discovery_2016,currie_matryoshka_2016}. Interestingly, our higher quality $Ls$ dataset collected on 2014-Apr-12 (start of night) hints at a point source-like object at the leading tip of the arc or spiral feature, perhaps due to a dust clump. Though this point source-like feature could be the result of the KLIP reduction, we encourage follow-up observations to confirm or rule out this peculiar feature. 

The \hd{} system is well-known for its large two outer rings; the outermost at $\sim400$ au and the second ring at $\sim200$ au. These outer disks also have confirmed spiral arms and arcs (e.g., \citealt{konishi_discovery_2016}; \citealt{biller_gemini_2015}), though these outer features curve in the opposing sense to that in the inner region. However, all spiral or arc features currently detected in this system are on the eastern side of the star. The existence of these spirals and arcs perhaps implicate gravitational perturbations either from the two known M dwarf companions or from ongoing planet formation within the disk system.

\subsection{Water Ice}
\label{sec:discuss-ice}

We carried out multifilter observations of the inner disk at high S/N at $L'$ (3.8 $\mu$m), $Ls$ (3.3 $\mu$m), and narrowband $Ice$ (3.1 $\mu$m). We attempted to observe the disk using a  $Ks$ (2.15 $\mu$m) filter, but no disk components were detected in these data. A dip in the brightness of scattered light at $3.1 \mu$m should be characteristic of the presence of water ice, as first noted by \cite{inoue_observational_2008}. The observed surface brightness in our $Ice$ band KLIP-RDI image might provide clues on the radial distribution of water ice. Figure \ref{fig:sb-profiles} shows that the $Ice$ KLIP-RDI profile curve for the south disk ansa is comparatively bright closer in towards the star which could be due to a decreasing population of icy grains at smaller stellar distances as seen in HD 100546 \citep{honda_detection_2009} and AB Aurigae \citep{betti_detection_2022} (discussed in the following subsection). This same evidence is not seen in the north ansa in our $Ice$ data, however, where the KLIP-RDI image profile indicates that region is dimmer compared to $Ls$ and $L'$ at all radial distances. These KLIP-RDI image profiles hint that the disk ansae detected at the $Ice$-band do not scatter less light in comparison to $Ls$ and $L'$ until about 40 au (based on this viewing geometry). Nonetheless, considering our extracted surface brightness and total disk flux densities from our modeling (see Section \ref{sec:results_sb}), our results suggest significant absorption due to icy grains at 3.1 $\mu$m.

We only considered silicates in our modeling as we found that constraining an appropriate mix of ice/silicate grains in our scattered light models was difficult, both due to the limited range of scattering angles probed and the limited wavelength range of the data. As such, we leave more concentrated efforts to test mixed grain models to a future study. \textbf{Despite the absence of a photometric measurement at $Ks$, we do not expect the flux density at this passband to be lower than our $Ice$ measurement. \cite{boccaletti_ground_2003} show that the scattered flux is larger at $1.6 \mu$m compared to $2.2 \mu$m in the outer two disk components. Moreover, \cite{inoue_observational_2008} show that $K$-band is always brighter relative to $Ice$-band using scattering models. Using our flux density measurements shown in Table \ref{tab:diskfm}, we compute a delta magnitude of $(Ice - L') = 0.51$ mag. Though we lack the $K$-band measurement to directly compare to the color-color diagram in \cite{inoue_observational_2008}'s Figure 5, we expect to be among the optically thin, icy grains classifications.}

A simple calculation to get the expected temperature of silicate grains around a star of $T_{\text{eff}} = 10000K$ and $L_{\odot} = 27.5$ revealed that the water ice sublimation zone terminates at around 6.5 to 12 au depending on the specific grain composition. This is clearly well within our IWA, but one other possibility is that water ice even beyond this estimated sublimation zone is undergoing photodesorption due to UV photons. For disks around Herbig Ae/Be stars, water ice at the surface is expected to have a short lifetime and will readily be destroyed by UV photons \citep{oka_effect_2012}. \textbf{If we are detecting water ice in the inner disk region of \hd{} ($\sim20$ to 70 au) and the disk material is not optically thin to UV flux responsible for quick photodesorption (e.g., \citealt{jonkheid_gas_2004}), there would need to be some way to explain how the ice is replenished over time. One way ice grains can migrate to the disk surface is through vertical mixing (see, e.g., \citealt{takeuchi_dust_2001}), perhaps due to gravitational perturbations on this disk, which is gas-rich \citep{di_folco_almanoema_2020}.} 

Detection of water ice has been done using this scattering deficit method for 3 other disk systems: HD 100546 \citep{honda_detection_2009}, HD 142527 \citep{honda_water_2016}, and AB Aurigae \citep{betti_detection_2022}. We follow with a brief overview of these 3 other icy disk systems.

\subsection{Comparing HD 141569 with Other Disks Showing Water Ice Absorption}
\label{sec:comparison_disks}

HD 100546, HD 142527, and AB Aurigae are all Herbig Ae/Be stars (as with HD 141569) whose moderate inclinations allow for comparison of the $\text{H}_2$O ice scattering feature in different regions at the disk surface.

\subsubsection{Optical Depth Effects}
\label{sec:optical_depth}

In the case of HD 100546 (Age$\geq10$ Myr; \citealt{van_den_ancker_hipparcos_1997}), \cite{honda_water_2016} found that the optical depth increases with distance from the star, suggesting a trend of increasing prevalence of icy grains further out from the star \textbf{which is consistent with our argument presented above.} This characteristic was echoed by the recent study of the AB Aurigae system (Age$= 4 \pm 1$ Myr) by \cite{betti_detection_2022} who also found a lower optical depth at smaller separations from \textbf{the star, as well} as a small asymmetry in optical depth with a higher value along the major axis. However, additional complexity is attributed to the HD 100546 disk considering strong asymmetry found in the absorption feature across the minor axis by \cite{honda_water_2016} which \cite{tazaki_water-ice_2021} interpreted as disk subsections possessing higher concentrations of icy grains. Differing optical depths along the major and minor axes were also observed in AB Aurigae by \cite{betti_detection_2022} although the asymmetry was much less pronounced.

\subsubsection{Photodesorption}
\label{sec:photodesorption}

For HD 142527, \cite{honda_detection_2009} performed radiative transfer modeling and concluded that this disk is rather ice rich, with an ice/silicate mass ratio of $\geq2.2$. However, \cite{tazaki_water-ice_2021} argued this result to be a consequence of the assumptions made to simplify the modeling such as isotropic scattering. Indeed, as pointed out by \cite{oka_effect_2012}, photodesorption of water ice by UV photon exposure at the scattering disk layer makes an ice-heavy composition difficult to maintain. \cite{tazaki_water-ice_2021} revisited radiative transfer modeling for the HD 142527 disk with added complexity to the scattering and disk geometry assumptions which yielded a lower ice/silicate mass ratio of 0.06 to 0.2 which is consistent with similar modeling efforts for AB Aurigae by \cite{betti_detection_2022} who derived an upper limit on the ice/silicate ratio of $\sim0.05$. Additionally, \cite{honda_subaruircs_2022} performed a spectropolarimetry analysis on HD 142527 and confirmed the $3.1 \mu$m scattering feature in the disk and derived an ice/silicate mass ratio generally consistent with \cite{tazaki_water-ice_2021}. There would be value in similar observations of other Herbig Ae/Be systems since the proximity of these ice/silicate mass ratios begs the question of whether this level of ice composition is a characteristic of them or entirely coincidental.

\subsection{Possible Detection of Thermal PAH Emission}
\label{sec:pahs}


We carried out observations at $Ls$ to probe for the $3.3 \mu$m PAH emission feature in the \hd{} disk system and an intriguing result from our forward modeling is the above-mentioned southwestern brightness feature in the best-fit disk models (Figure \ref{fig:master_diskfm_results}). This brightness enhancement is enigmatic since we assume the east side of the disk is closest to us (see, e.g., \citealt{bruzzone_imaging_2020}) and realistic dust grains tend to forward scatter. One explanation for excess emission from the far side of the disk is PAH emission. This is supported by the brightness of the feature relative to the rest of the disk at $Ls$ when compared to our other models in the adjacent continuum. PAH emission from stochastic, single-photon heating does not have the forward/backward asymmetry that results from scattering \citep{draine_infrared_2001}. PAH emission may be visible in the back scattering regions as well as beyond the peak dust density, providing the PAH grains are illuminated directly by stellar UV and are optically thin to our line of sight.

We attempted a simple experiment to try to constrain the flux density associated with this peculiar brightness feature by generating a $Ls$ disk model with stellar offsets set to zero and subtracting it from the best $Ls$ model displayed in Figure \ref{fig:master_diskfm_results}. We then integrated the residuals in only this southwest region using an irregular aperture to obtain a flux density estimate of $\sim 2.7$ mJy. Similarly, we repeated these steps, but generated our test model using the parameter solutions for the best $Ice$ model, but retaining the flux scaling parameter for $Ls$ which yielded a flux density estimate of $\sim 4.6$ mJy. Thus if we assume the increase in brightness in the southwestern region is purely due to $3.3 \mu$m emission from PAHs, we detect an associated flux density of around 3 -- 5 mJy.

We can compare this PAH emission to what might be expected from the disk.  First, \cite{geers_spatially_2007} found a spectroscopic upper limit to the flux density from $3.3 \mu$m PAH emission of $6.9 \times 10^{-16}$ W m$^{-2}$ for the \hd{} system. From \cite{castro_akari_2014}, we can use a typical line width of $0.03 \mu$m for the $3.3 \mu$m feature to estimate the total flux density upper limit at 83 mJy. However, we note that this assumes the emitting region was captured within the  $0\farcsec6$ wide slit used by \citet{geers_spatially_2007}. Second, the PAH lines at $6.26 \mu$m, $7.98 \mu$m, and $11.28 \mu$m were well-measured in Spitzer spectra \citep{sloan_mid-infrared_2005}, with flux ratios reported as $F_{6.2}/F_{7.9} = 0.4$ and $F_{11.2}/F_{7.9} = 0.04$. We can compare these to the expected flux ratios from PAHs with different sizes and ionization states \citep{draine_excitation_2021}. Comparing to \cite{draine_excitation_2021}'s Figure 21, \hd{} has lowest ratios for $F_{6.2}/F_{7.7} = 0.28$ and $F_{11.3}/F_{7.7} = 0.13$, which indicates small PAHs in a high UV flux. This implies a flux ratio for $F_{3.3}/F_{7.7}$ of 0.1 -- 0.2, which in turn gives a flux density prediction of the 3.3 $\mu$m line of 1.7 -- $3.4 \times 10^{-15}$ W m$^{-2}$. We also downloaded a model dust emission spectrum from \cite{draine_infrared_2007} corresponding to $U = 3 \times 10^5$\symfootnote{Dimensionless quantity denoting the relative starlight intensity, where $U = 1$ describes the intensity within the local interstellar medium.}, $R_V \sim 3.1$ (typical extinction within the Milky Way), and PAH mass fraction $q_{\text{PAH}} = 4.48 \%$. We subtracted the continuum from small grains before calculating $F_{3.3}/F_{6.6} = 0.27$, which then predicts a $3.3 \mu$m flux of $1.85 \times 10^{-15}$ W m$^{-2}$ or 218 mJy using the line width from before. \textbf{Our estimates are all well below these limits and may point to a stronger PAH abundance interior to our IWA to explain the observationally-inferred value. This idea is supported by the recent detection of a PAH-dominated dusty ring at $\sim 1$ au by GRAVITY \citep{gravity_collaboration_gravity_2021}.}

It has been proposed (\citealt{seok_polycyclic_2017}, \citealt{maaskant_polycyclic_2014}, \citealt{acke_spitzers_2010}, \citealt{keller_pah_2008}) that PAHs might trace the geometry of the disk gas. As discussed in \cite{di_folco_almanoema_2020}, the gas around \hd{} has a flared geometry, which would enable UV photon exposure of the upper disk atmosphere. Additionally, results from \cite{keller_pah_2008} indicate that \hd{} possesses the most ionized PAH spectrum compared to the other systems surveyed in that study (see their Figure 6). Also suggested in \cite{keller_pah_2008} is the notion that icy mantles on large grains could release PAHs given optical-UV photon exposure, which is consistent with our probable ice detection (see Section \ref{sec:discuss-ice} above). Recent studies using ALMA by \cite{di_folco_almanoema_2020} and \cite{white_alma_2016} found a $^{12}$CO $J = 2 \rightarrow 3$ emission asymmetry, with a CO peak in the western region of the disk that is colocated with both the brightness asymmetry we report here as well as the peak dust density region found in \cite{singh_revealing_2021}. \textbf{This correspondence between the dust and gas in a restricted location of the disk supports the idea that water ice is resupplied at the surface layer by vertical mixing through drag forces even if the disk is optically thin to the starlight (e.g., \citealt{takeuchi_dust_2001};\citealt{lyra_formation_2013})}. Specifically, collisions between \textbf{icy planetesimals} could release a substantial amount of gas, as has been similarly hypothesized for the $\beta$-Pictoris debris disk \citep{dent_molecular_2014}. \textbf{From the above points, it is possible that our modeling results at $Ls$ and $L'$ hint at a complicated spatial distribution of the $3.3 \mu$m PAH emission feature. From this, we would also expect a brightness enhancement in the southwestern region in the $Ice$-band image from a deficit of water ice. Figure \ref{fig:master_diskfm_results} reveals a significant amount of under-fit flux directly south of the star (bottom right panel), but this location is neither sufficiently far nor close to the peculiar bright region seen in the $Ls$ model to make any definitive conclusions. There would be value in additional observations to test this hypothesis.} We encourage more work on constraining the distribution of PAHs in more disk systems to elucidate what exactly they can tell us about the disk environment and ultimately inform us on the stages leading up to early planet formation.

\section{Summary \& Future Work} 
\label{sec:summary}

We observed \hd{} using the MagAO+Clio instrument at $L'$ ($3.8 \mu$m) on UT 2014-04-09, $Ls$ ($3.3 \mu$m) on UT 2014-04-11, and narrowband $Ice$ ($3.1 \mu$m) on UT 2015-05-29. We performed PSF subtraction using the KLIP-RDI algorithm to detect the inner disk component at S/N of $\sim10$ at $L'$, $\sim8$ at $Ls$, and $\sim10$ at $Ice$. We then used our KLIP-RDI images to accomplish simple scattered light disk model fitting using the DiskFM pipeline in \texttt{pyKLIP}. Through our observations we confirmed complex geometrical features such as non-axisymmetric brightness and an arc or spiral feature. We detected a potential point source-like feature at the aforementioned arc's terminus to which we encourage follow-up observations for confirmation. The best-fit $Ls$ and $L'$ models preferentially exhibit peak brightness in the southwest region which could imply the detection of $3.3 \mu$m emission due to PAHs. Additionally, a significantly lower surface brightness for our best-fit $Ice$ model compared to our best-fit $L'$ model indicates probable detection of water ice in this inner disk region.

Overarching goals for this study included further constraint on the geometrical morphology of the inner disk by forward modeling a simple, scattered light disk model and probing the minimum grain size using a realistic SPF calculated from ADPs \citep{arnold_stumbling_2022}. Results of these efforts are largely consistent with previous studies with the exception of the disk inclination $i \approx 48^{\circ}$, which is about $10^{\circ}$ less than previous literature values, and a surprisingly high flux from the western half of the disk, despite it being the back-side. One explanation for this western brightness enhancement is the $3.3 \mu$m PAH emission feature. This would be co-located with a CO gas enhancement and dust density enhancements seen in other scattered light images (\citealt{white_alma_2016}; \citealt{di_folco_almanoema_2020}; \citealt{singh_revealing_2021}). A more complex model is outside the scope of this paper, but may provide answers on which morphological features, if any, are better fit with a combination of scattered light and PAH emission. Furthermore, since our grain size modeling suggests the majority of scattered light detected is from sub-micron grains, it is apparent that the west side of the inner disk presents an ideal opportunity to leverage modeling of the dynamical interactions between the dust and gas, which we encourage as an avenue for future work.

Our analysis of the surface brightness estimates from our best-fitting models suggest significant scattered light absorption at $Ice$ $(3.1 \mu$m) perhaps due to icy grains permeating the surface layers of the inner disk. Unfortunately, observations using a bordering $Ks$ filter failed to detect the disk. Additional total intensity observations at $Ks$ or bluer to probe the surface-brightness spectrum are needed to confirm these results and/or characterize the amount of ice. If water ice is present in the upper scattering layers of the inner disk, icy grains would need to be resupplied as their lifetimes in these areas are short due to photodesorption via UV photons exposure. Previously mentioned features such as the spiral or arc-like feature as well as the azimuthal variation of dust and gas hints at gravitational perturbations affecting this inner disk system. To what extent this encourages vertical mixing to cause icy grains to migrate from the midplane to upper disk atmosphere is an analysis saved for a later study.

\section*{Acknowledgements}

We are grateful to the anonymous reviewers whose thoughtful feedback vastly improved the quality of this manuscript.

\ We thank Prof. Maxwell Millar-Blanchaer for helpful advice during critical modeling steps. Suggestions from Justin Hom benefitted our modeling strategies and overall quality of this manuscript markedly. We would like to give thanks to Joseph Long who offered helpful comments at various stages of this analysis.

\ This work has made use of data from the European Space Agency (ESA) mission
{\it Gaia} (\url{https://www.cosmos.esa.int/gaia}), processed by the {\it Gaia}
Data Processing and Analysis Consortium (DPAC,
\url{https://www.cosmos.esa.int/web/gaia/dpac/consortium}). Funding for the DPAC
has been provided by national institutions, in particular the institutions
participating in the {\it Gaia} Multilateral Agreement.

\ This material is based on work supported by the National Science Foundation Graduate Research Fellowship Program under grant No. 2020303693. Any opinions, findings, and conclusions or recommendations expressed in this material are those of the author(s) and do not necessarily reflect the views of the National Science Foundation.



\ This publication makes use of data products from the Two Micron All Sky Survey, which is a joint project of the University of Massachusetts and the Infrared Processing and Analysis Center/California Institute of Technology, funded by the National Aeronautics and Space Administration and the National Science Foundation.

\vspace{5mm}

\appendix

\section{MCMC Modeling Corner Plots}
\label{sec:appendixA}

\renewcommand\thefigure{\thesection\arabic{figure}}
\setcounter{figure}{0}
\renewcommand\thetable{\thesection.\arabic{table}}
\setcounter{table}{0}

\begin{figure}
    \centering
    \includegraphics[width=0.75\linewidth]{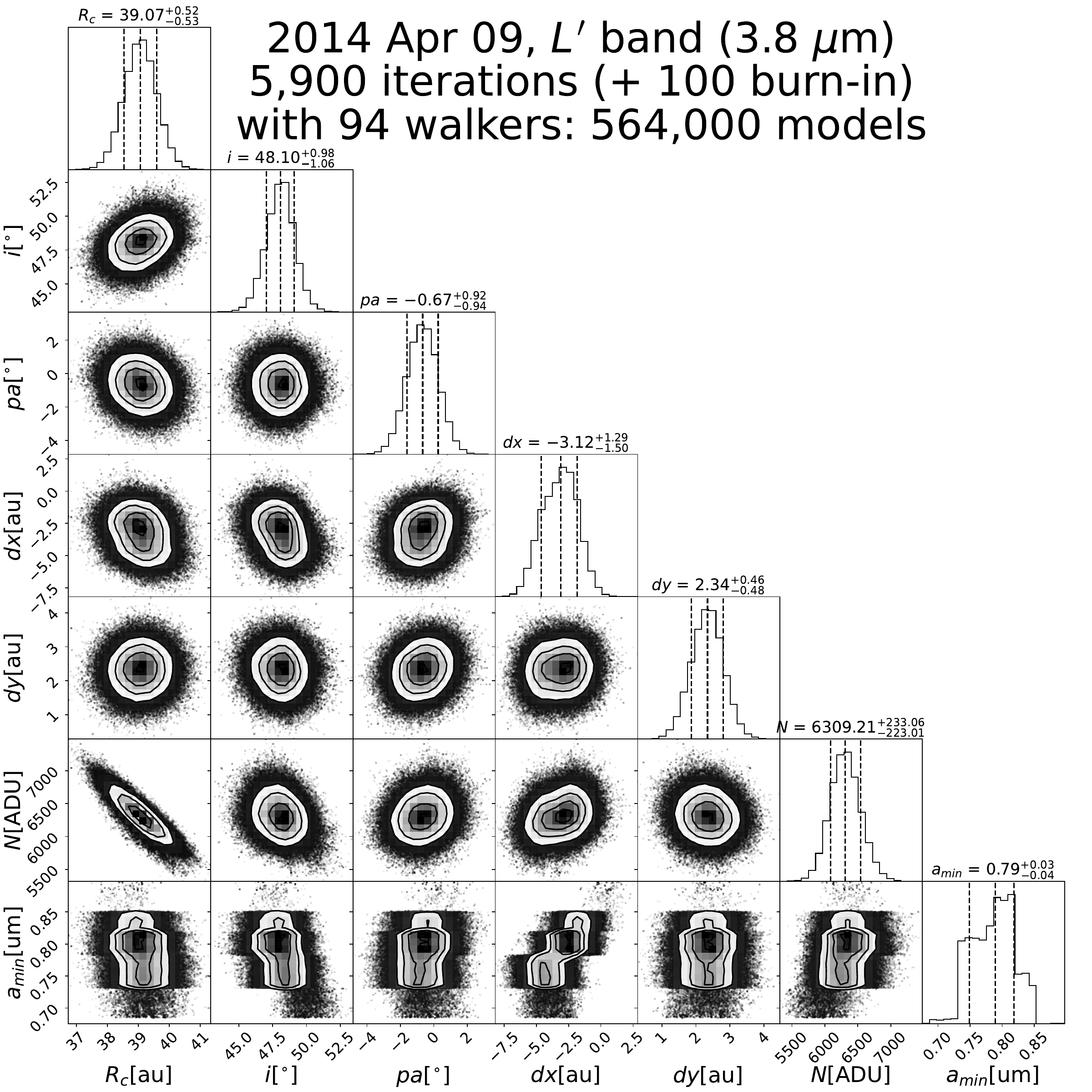}
    \caption{MCMC modeling results using DiskFM for the $L' (3.8 \mu m)$ band. Histograms on the main diagonal show the posterior probability distributions for each fitted parameter marginalized over all other fitted parameters where the dashed vertical lines show the 16th, 50th, and 84th percentiles. The off-diagonal scatter plots show the joint probability distributions overlaid with contours also representative of the aforementioned percentiles.}
    \label{fig:corner_38}
\end{figure}

\begin{figure}
    \centering
    \includegraphics[width=0.75\linewidth]{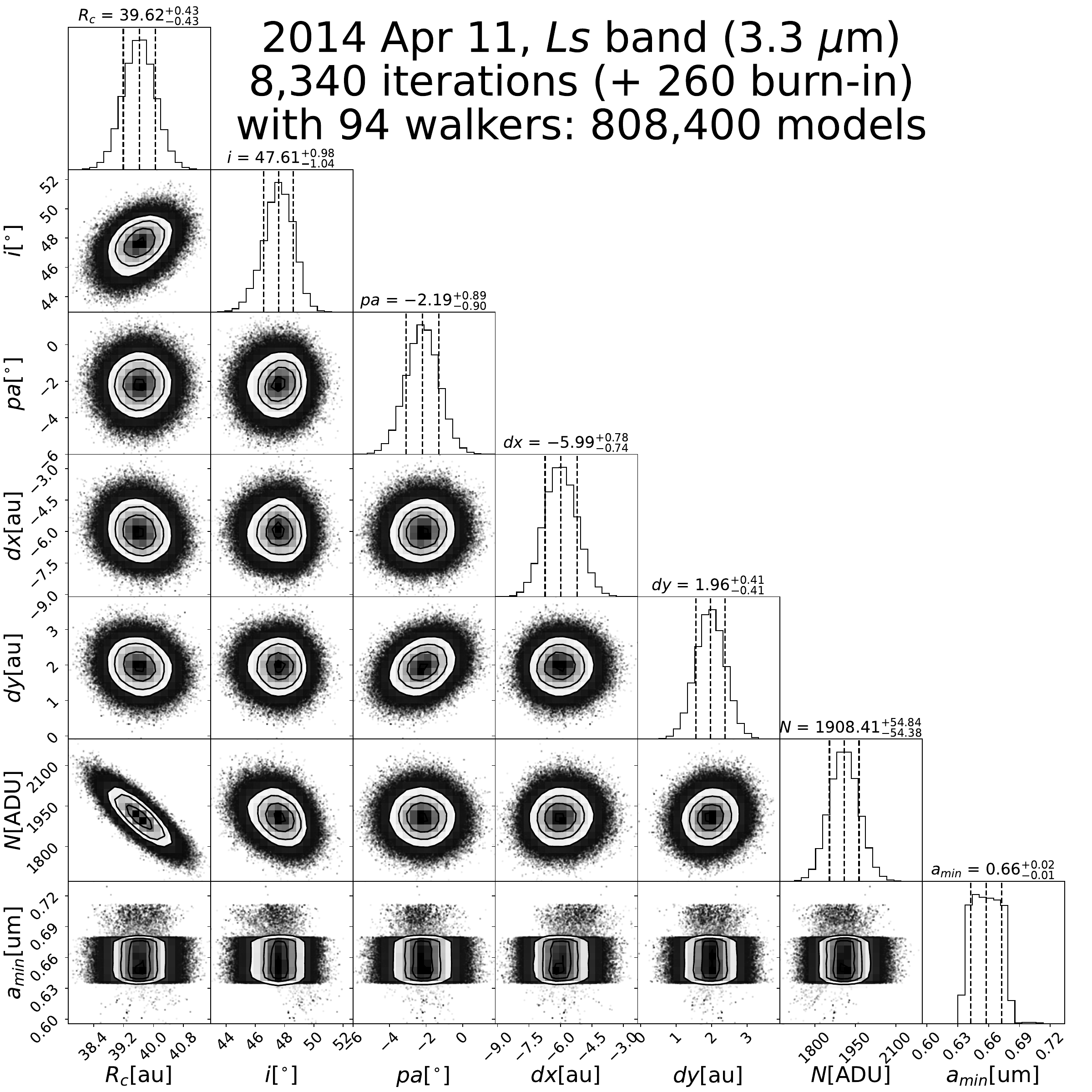}
    \caption{Same as Figure \ref{fig:corner_38}, but for the $Ls (3.3 \mu m$) band data.}
    \label{fig:corner_33}
\end{figure}

\begin{figure}
    \centering
    \includegraphics[width=0.75\linewidth]{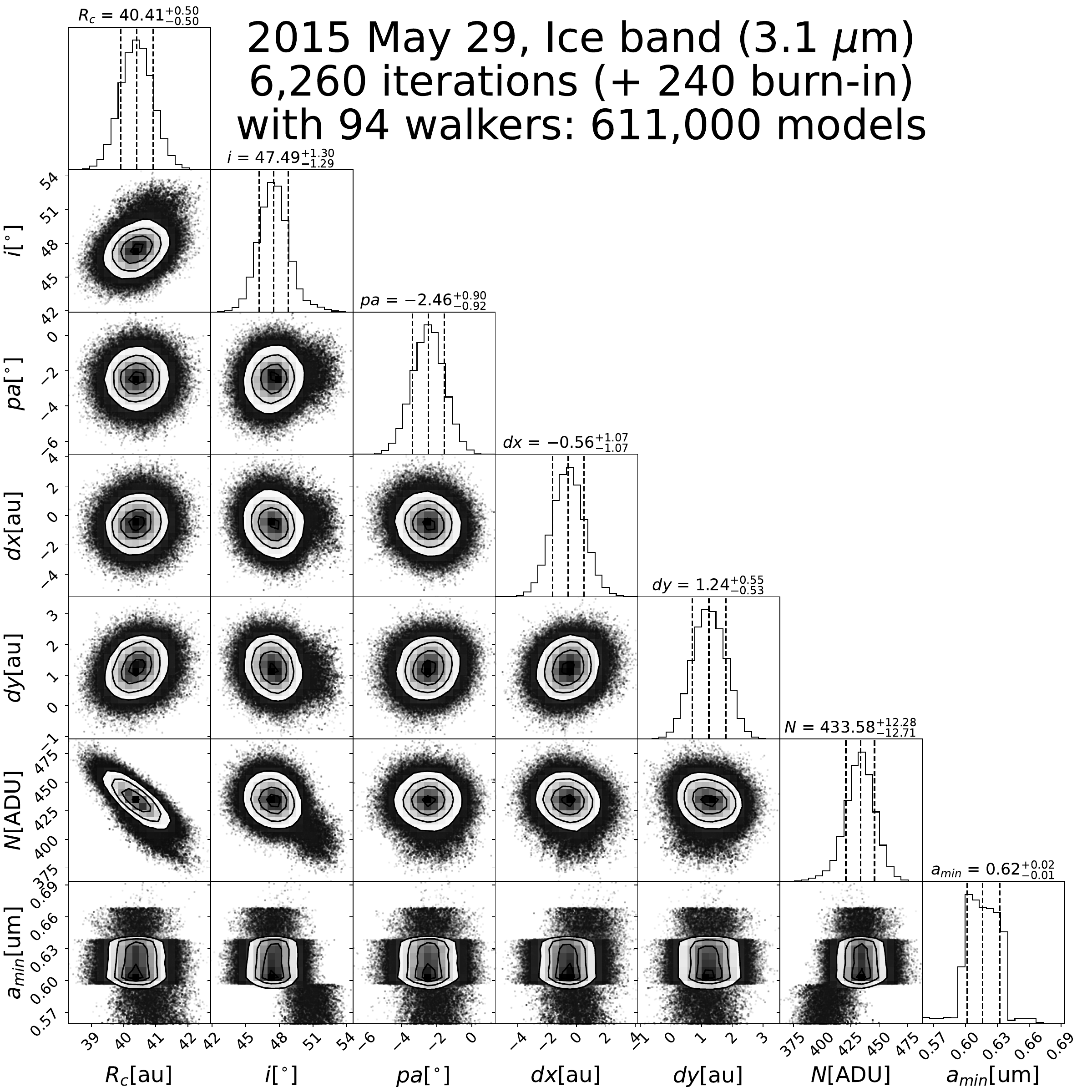}
    \caption{Same as Figure \ref{fig:corner_38}, but for the $Ice (3.1 \mu m$) band data.}
    \label{fig:corner_31}
\end{figure}

Figures \ref{fig:corner_38}, \ref{fig:corner_33}, and \ref{fig:corner_31} show corner plots illustrating the output sample posterior distributions from our MCMC runs. The values shown above the histogram plots correspond to those shown in Table \ref{tab:diskfm}
. Fitted parameters include the disk critical radius $R_c$, inclination $i$, position angle $pa$, x-offset $dx$, y-offset $dy$, a flux scaling parameter $N$, and minimum grain size $a_{\text{min}}$ (see Section \ref{sec:model_desc} for further details on the parameters utilized in our modeling). We ran all MCMC runs until the number of iterations exceeded 50 times the maximum autocorrelation time of the walker chains (computed using included functionality of the \pkg{emcee} package; \citealt{foreman-mackey_emcee_2013}) and discarded the number of initial steps equal twice the autocorrelation time as recommended by the \pkg{emcee} documentation \footnote{\url{https://emcee.readthedocs.io/en/stable/user/autocorr/}}. As a sanity check, visual inspection of the traces of the walker paths showed that the walkers had stabilized well outside of both of these criteria.

\clearpage

\clearpage

\centering
\clearpage

\bibliography{clio_HD141569A_references}{}
\bibliographystyle{aasjournal}

\end{document}